\documentclass[11pt,a4paper]{article}

\usepackage[a4paper,margin=1.0in]{geometry}
\usepackage{graphicx}
\usepackage[percent]{overpic}
\usepackage{amsmath,amssymb}
\usepackage{appendix}
\usepackage[round,authoryear]{natbib}
\usepackage{xcolor}
\usepackage{tcolorbox}
\usepackage{multicol}
\usepackage{multirow}
\usepackage{booktabs}
\usepackage[font=small,labelfont=bf]{caption}
\usepackage{microtype}
\usepackage{authblk}
\usepackage[hidelinks]{hyperref}
\usepackage{algorithm}
\usepackage{algpseudocode}

\tcbuselibrary{listings}

\title{\bfseries A global mass-preserving numerical method for low-Mach-number real-gas flows in closed systems}

\author[1]{Sanath Kotturshettar\thanks{Corresponding author: \texttt{S.B.Kotturshettar@tudelft.nl}}}
\author[1,2]{Gonçalo Alcobia}
\author[3]{Pietro Carlo Boldini}
\author[1]{Rene Pecnik}
\author[1]{Pedro Costa}

\affil[1]{%
Process \& Energy Department\\
Faculty of Mechanical Engineering\\
Delft University of Technology\\
Leeghwaterstraat 39, 2628~CB Delft, The Netherlands
}

\affil[2]{%
CEFT, ALiCE\\
Faculty of Engineering\\ University of Porto\\
Rua Dr.\ Roberto Frias, s/n 4200-465 Porto, Portugal
}

\affil[3]{%
Department of Flow Physics and Technology\\
Faculty of Aerospace Engineering\\
Delft University of Technology\\
Kluyverweg 1, 2629~HS Delft, The Netherlands
}

\date{}

\begin{document}

\maketitle

\begin{abstract}
\noindent

A mass-preserving low-Mach-number framework is proposed for closed-system real-fluid flows governed by general nonlinear equations of state.
The formulation enforces consistency between the spatially uniform thermodynamic pressure, the equation of state, and global mass conservation. 
Moreover, the numerical algorithm employs a segregated strategy in which the thermodynamic state is updated before the momentum equations, and the velocity field is advanced using a pressure-correction method. 
This approach enables an efficient solution procedure by decoupling the thermodynamic and momentum updates and retaining the use of FFT-based solvers for the pressure correction. The resulting formulation is implemented with second-order spatial accuracy. The implementation is first verified using the method of manufactured solutions, in which the thermodynamic state is prescribed through analytical density and thermodynamic-pressure fields, enabling verification of the nonlinear equation of state, thermodynamic-pressure evolution, and the low–Mach-number divergence constraint. The framework is subsequently validated against benchmark laminar and turbulent flows for both ideal and real fluids, particularly transcritical CO$_2$ channel flow, demonstrating its accuracy and robustness in the presence of strong thermodynamic nonlinearities.

\end{abstract}

\section{Introduction}\label{sec:intro}
Turbulent flows with large temperature variations occur in a wide range of natural and industrial applications, including heat exchangers, turbomachinery, combustion systems, and atmospheric and oceanic mixing processes \citep{zonta_stably_2018}. Many of these are characteristically low-speed flows (relative to the speed of sound), with negligible acoustic effects in the flow dynamics. In these cases, using a fully compressible flow solver is computationally inefficient: resolving these very small acoustic time scales requires timesteps that become vanishingly small, relative to the convective time scale, in the limit of zero Mach number. A common strategy to overcome this limitation is to adopt formulations of the governing equations that remove the acoustic effects \citep{Majda1985TheCombustion,Almgren_2006,Peeters_Pecnik}.

The Oberbeck--Boussinesq approximation represents one such simplification and is valid when temperature variations are small and the resulting density variations are negligible \citep{oberbeck1879ueber,landau1987fluid, kundu2024fluid}. In this regime, the flow can be treated as incompressible, with density variations only appearing in the buoyancy term of the momentum equations. However, the approximation breaks down when temperature variations are large enough to cause significant density fluctuations (typically about $30\ \mathrm{K}$ in air; see \cite{Gray1976TheVO}). Significant density variations may also arise even for small temperature differences, for example in supercritical fluids near the critical point \citep{yoo2013turbulent}. For such flows, if the fluid velocity remains small relative to the speed of sound, the governing equations may instead be simplified using the low-Mach-number approximation. 

The low-Mach-number approximation is derived by expressing the dependent variables, like pressure, as asymptotic expansions in powers of the Mach number and substituting these expansions into the compressible Navier–Stokes equations \citep{Majda1985TheCombustion, mcmurtry_direct_1986}. The resulting set of equations, in the limit of vanishing Mach number, features a zeroth-order (\emph{thermodynamic}) pressure that is spatially uniform, and a first-order (\emph{hydrodynamic}) pressure that drives the fluid motion. This decomposition separates the thermodynamic and hydrodynamic components of pressure. Because only the hydrodynamic pressure enters the momentum equations, acoustic modes are absent from the resulting formulation. As a result, the time step is no longer limited by acoustic-wave propagation, whose speed becomes very large relative to the flow velocity in low–Mach–number conditions, while temperature-induced density variations and their dynamical effects on the flow are still retained.

The hydrodynamic pressure governs the local flow dynamics and is determined by enforcing the velocity-divergence constraint. In low-Mach-number flows, the velocity divergence is determined by the local volumetric expansion or contraction of the fluid due to heat transfer and by the temporal evolution of the thermodynamic pressure. The thermodynamic pressure, in contrast, determines the system's global thermodynamic state and is obtained from a global thermodynamic constraint. This constraint depends on whether the system is open or closed: For open systems, the thermodynamic pressure remains constant and equals its prescribed initial value \citep{Cook1996DirectComputer}; in closed systems, the thermodynamic pressure evolves in time and must be determined so that the total mass of the system remains constant \citep{nicoud_conservative_2000}.

A consistent evaluation of the thermodynamic pressure is essential for accurately representing the system's thermodynamic state. The equation of state relates density to the thermodynamic pressure and temperature and plays a central role in low-Mach-number formulations. Such formulations, closed with the ideal-gas (IG) equation of state, have been extensively used to study flows with strong property variations, particularly in the context of reactive flows \citep{Yu2012AnSolver,Motheau2016AAccuracy,Nonaka2018AIntegration,Bartholomew2019ACurrents}, in both open and closed systems. Furthermore, the linear dependence of density on pressure in the ideal-gas equation of state enables several simplifications, including an analytical update of the thermodynamic pressure in closed systems \citep{nicoud_dns_1999}.

Many practical applications, however, involve fluids operating in regimes where non-ideal effects are important, such as supercritical fluids near the critical point \citep{guardone2024nonideal}. Here, it becomes essential to account for the real-gas behavior through a non-linear equation of state. For example, \citet{accary2009turbulent} performed three-dimensional simulations of Rayleigh–Bénard convection in a supercritical Van der Waals fluid. Similarly, \citet{Battista2014TurbulentNumber} developed a low-Mach-number formulation for non-ideal fluids and applied it to a coaxial jet of a slightly supercritical Van der Waals fluid. Since the latter configuration was an open system, the thermodynamic pressure remained constant throughout the simulation. However, in closed systems, evaluating the thermodynamic pressure becomes considerably more challenging because it must satisfy the global mass-conservation constraint alongside a non–linear equation of state. \citet{guo2022structure} and \citet{wan2025effects} performed direct numerical simulations (DNS) of turbulent channel flows at transcritical\footnote{A transcritical process follows a thermodynamic path across the Widom line, through the pseudo-boiling region, from a liquid-like state to a gas-like state.} conditions using fully compressible and low-Mach-number formulations, respectively. Although the channel geometry was periodic and bounded by impermeable walls, the thermodynamic pressure was prescribed in both cases, thereby treating the system as formally open. Consequently, the integral of the density obtained from the equation of state was not explicitly constrained, allowing a small formal mass loss during the thermal transient \citep{Nemati2015MeanPressure}. A constant thermodynamic pressure has also been used successfully in DNS of variable-property ideal-gas channel flows, including the thermally balanced configurations of \citet{Patel2015SemiLocal}. In such balanced or statistically stationary channel flows, the resulting global discrepancy can remain modest, and the corresponding DNS data therefore remain valuable benchmarks. The distinction is nevertheless important for general closed real-fluid systems: the present method eliminates this uncontrolled approximation by enforcing total mass exactly while allowing the thermodynamic pressure to evolve.

Accordingly, a mass-preserving numerical method based on the low-Mach-number formulation is developed for real-fluid flows in closed systems. An iterative procedure is introduced to determine the thermodynamic pressure while enforcing global mass conservation. The proposed approach exploits thermodynamic quantities readily available from equations of state or lookup tables and therefore does not depend on a particular analytical form of the thermodynamic model. As a result, the framework can be coupled directly to tabulated thermodynamic data and naturally extended to arbitrary real fluids within a stable single-phase regime, including supercritical fluids that exhibit strongly nonlinear thermodynamic behavior.

For the numerical solution procedure, the present work adopts a segregated algorithm, a commonly used approach for low-Mach-number flows \citep{Knikker2011AFlows}. The algorithm consists of two stages. First, the thermodynamic state is advanced by solving the temperature equation and subsequently evaluating the thermodynamic pressure. The updated thermodynamic state is then used to compute the remaining thermophysical properties, including density. Second, the momentum equations are advanced using the updated thermodynamic fields.

The momentum equations are integrated using a pressure-correction method \citep{amsden1970simplified}, analogous to those employed in incompressible flow solvers. 
However, in these flows, the pressure-correction step leads to a variable-coefficient Poisson equation due to spatial variations in density. To avoid the computational expense of solving such equations, a pressure-splitting strategy is employed \citep{Dong2012ARatios, Dodd2014AFlows}, which recasts the problem as a constant-coefficient Poisson equation. This approach has previously been applied to low-Mach-number systems by \citet{Motheau2016AAccuracy} and \citet{Demou2019AFlows}.

The implementation is verified using the method of manufactured solutions, with particular emphasis on the nonlinear equation of state, the evolution of the thermodynamic pressure, and the enforcement of the velocity-divergence constraint. The verification study also establishes the formal order of accuracy of the solver. The framework is subsequently validated against the laminar differentially heated cavity benchmark of \citet{LeQuere2005ModellingSolutions} and \citet{Demou2019AFlows} for an ideal gas with large temperature differences, and against the turbulent ideal-gas channel-flow simulations of \citet{nicoud_dns_1999}. The differentially heated cavity with Van der Waals fluid is also simulated to compare against the ideal gas results. Finally, turbulent channel flows at transcritical conditions are simulated to demonstrate the applicability of the proposed framework to real-fluid flows, and the results are compared with existing DNS data from \citet{wan2025effects}.

The remainder of this paper is organized as follows. Sections~\ref{sec:gov_eq} and \ref{sec:num_me} present the governing equations and numerical methodology. Verification and validation results are presented in sections~\ref{sec:MMS} and \ref{sec:results}, respectively. The application to turbulent channel flows at transcritical conditions is described in section~\ref{sec:TCF}. Finally, the main conclusions are summarized in section~\ref{sec:conclusion}.

\section{Governing equations}\label{sec:gov_eq}

The underlying idea of the low–Mach–number approximation is the removal of acoustic effects from the fully compressible Navier–Stokes equations. To achieve this, following \citet{Majda1985TheCombustion}, the dependent variables are expressed as asymptotic expansions in powers of $\gamma_{\mathrm{p\upsilon}} Ma^2$, where $\gamma_{\mathrm{p\upsilon}}$ (see Eq.\eqref{eq:isent_gamma}) denotes the isentropic exponent \citep{Nederstigt2023GeneralisedThermodynamics}. For an ideal gas, $\gamma_{\mathrm{p\upsilon}}$ reduces to $\gamma$, the ratio of specific heats. The pressure can therefore be expanded as
\begin{equation}
    p = p^{(0)} + (\gamma_{\mathrm{p\upsilon}} Ma^2)p^{(1)}+ (\gamma_{\mathrm{p\upsilon}} Ma^2)^2p^{(2)}+\cdots,
\end{equation} 
with analogous expansions applied to the velocity, temperature and density. Substitution of these expansions into the fully compressible governing equations, followed by the collection of terms of equal order in $\gamma_{\mathrm{p\upsilon}} Ma^2$, yields the low–Mach–number formulation. A detailed derivation of this approximation is provided in Appendix~\ref{sec:LMDerivation}.

The derivation presented in Appendix~\ref{sec:LMDerivation} shows that, within the low–Mach–number formulation, the leading-order pressure is spatially uniform, whereas the first-order pressure varies in space. The pressure field can therefore be written as
\begin{equation}
    p(\mathbf{x},t) = p_0(t) + p_1(\mathbf{x},t),
\end{equation}
where $p_0$ is the \emph{thermodynamic} pressure, $p_1$ is the \emph{hydrodynamic} pressure, and the factor $\gamma_{\mathrm{p\upsilon}} Ma^2$ is absorbed into the definition of $p_1(\mathbf{x},t)$, such that $p_1(\mathbf{x},t) = (\gamma_{\mathrm{p\upsilon}} Ma^2)p^{(1)}$.

This decomposition is intuitive when considering the physics of a flow at vanishing Mach: In this limit, the speed of sound is so high relative to the characteristic flow velocity that perturbations in the thermodynamic pressure ($p_0$) rapidly propagate throughout the domain. Consequently, this quantity is spatially invariant and may only vary in time. Furthermore, because only pressure gradients appear in the momentum equations, thermodynamic pressure does not directly contribute to the momentum balance, and acoustic speeds do not impose a restrictive stability constraint on the time step. In contrast, the hydrodynamic pressure ($p_1/p_0\sim \mathcal{O}(\gamma_{\mathrm{p\upsilon}} Ma^2)$) varies in both space and time and determines the spatial pressure gradients driving the flow. 

Applying the low-Mach number approximation to the system of fully compressible conservation equations of mass, momentum, and sensible enthalpy yields the following set of governing equations for low–Mach–number flows:

\begin{subequations}\label{eq:governing_system}
\begin{equation}\label{eq:mass}
    \frac{\partial \rho}{\partial t} + \frac{\partial \rho u_j}{\partial x_j} = 0;
\end{equation} 

\begin{equation}\label{eq:momentum}
    \frac{\partial (\rho u_i)}{\partial t} + \frac{\partial }{\partial x_j}({\rho u_i u_j}) = -\frac{\partial p_1}{\partial x_i} + \frac{\partial \tau_{ij}}{\partial x_j} + \rho f_i;
\end{equation}

\begin{equation}\label{eq:enthalpy}
     \frac{ \partial (\rho h) }{\partial t} + \frac{\partial }{\partial x_j}({\rho u_j h}) = \frac{\partial }{\partial x_j}\left(\lambda \frac{\partial T}{\partial x_j}\right)  + \dot{\mathcal{Q}}  + \frac{\mathrm{d} p_0}{\mathrm{d} t},  
\end{equation}
\end{subequations}
where $\tau_{ij} = \mu (\partial u_i/ \partial x_j + \partial u_j/ \partial x_i - 2/3\ \partial u_k/\partial x_k\ \delta_{i,j})$, $f_i$ is any body force and $\dot{\mathcal{Q}}$ is the volumetric heat source. 

Several equivalent formulations of the energy conservation equation (Eq.\eqref{eq:enthalpy}) exist \citep{poinsot2005theoretical}. In the present work, the evolution equation for the sensible enthalpy ($h$) is employed, following the approach adopted in \citet{nicoud_dns_1999, Najm1998AChemistry, Nonaka2018AIntegration}. By contrast, \citet{Battista2014TurbulentNumber}, for example, formulated the low–Mach–number equations using conservation of internal energy.

In addition to the conservation equations, the system also includes an equation of state that describes the relationship between thermodynamic pressure, density, and temperature. To allow for real gas behavior, a generic equation of state is prescribed for a single-component, single–phase fluid as
\begin{equation}\label{eq:eos}
    p_0 = p(\rho, T).
\end{equation}
Similarly, other thermophysical properties such as thermal conductivity ($\lambda$), isobaric heat capacity ($C_p$), and dynamic viscosity ($\mu$) are specified through appropriate constitutive relations that are consistent with the chosen equation of state. Note that, within the low–Mach–number approximation, the hydrodynamic pressure, $p_1$, contributes to the density only at $\mathcal{O}(\gamma_{p\upsilon} Ma^2)$ and is therefore consistently neglected.

Since the equation of state and other properties are prescribed as functions of temperature, it is convenient to rewrite Eq.~\eqref{eq:enthalpy} in terms of temperature, using thermodynamic and Maxwell relations as described in Appendix~\ref{sec:enth2temp}. The resulting enthalpy transport equation in terms of temperature is
\begin{equation}\label{eq:cons_T}
    \rho C_p\frac{ \partial  T }{\partial t} + \rho C_pu_j\frac{\partial T}{\partial x_j} = \frac{\partial }{\partial x_j}\left(\lambda \frac{\partial T}{\partial x_j}\right)  + \dot{\mathcal{Q}}  + \beta T\frac{\mathrm{d} p_0}{\mathrm{d} t},
\end{equation}
where the specific heat capacity at constant pressure is $C_p = \left.\partial h/\partial T\right|_p$, and the coefficient of thermal expansion is $\beta = -(1/\rho)\left.\partial \rho/\partial T\right|_p$.
The appearance of the factor $\beta T$, which reduces to unity for an ideal gas, multiplying ${\mathrm{d} p_0}/{\mathrm{d} t}$ is one of the consequences of using real gas thermophysical properties.

Additionally, the mass conservation equation (Eq.~\eqref{eq:mass}) is reworked to obtain an expression for the divergence of velocity \citep{nicoud_dns_1999}. This is crucial to the pressure correction algorithm implemented in the present work. Within this framework, an intermediate velocity field is projected onto a space that satisfies a prescribed divergence constraint. Unlike incompressible flows, the velocity field is not divergence-free; instead, the divergence is given by
\begin{equation}\label{eq:mass_2}
     \frac{1}{\rho} \frac{\mathrm{D} \rho}{\mathrm{D} t} = -\frac{\partial u_i}{\partial x_i},
\end{equation}
which follows directly from rewriting Eq.~\eqref{eq:mass}. Since the density may be expressed as a function of the spatially uniform thermodynamic pressure and temperature ($\rho(p_0, T)$), Eq.~\eqref{eq:mass_2} can be further expressed as
\begin{equation}\label{eq:drhodt}
    \frac{1}{\rho} \frac{\mathrm{D} \rho}{\mathrm{D} t} = -\beta \frac{\mathrm{D} T}{\mathrm{D} t} + \chi \frac{\mathrm{d} p_0}{\mathrm{d} t} = -\frac{\partial u_i}{\partial x_i},
\end{equation}
where $\beta$ is the coefficient of thermal expansion defined above and $\chi = (1/\rho)\left.\partial \rho/\partial p\right|_T$ is the isothermal compressibility.
It is worth emphasizing that these thermophysical properties are functions of thermodynamic pressure and temperature, and can be obtained from analytical expressions of the equation of state or tabulated data.

Substituting Eq.~\eqref{eq:cons_T} into Eq.~\eqref{eq:drhodt} and simplifying yields
\begin{equation}\label{eq:divU_simplified}
     \frac{\partial u_i}{\partial x_i} = \frac{\beta}{\rho C_p} \left( \frac{\partial }{\partial x_i}\left(\lambda \frac{\partial T}{\partial x_i}\right)  + \dot{\mathcal{Q}} \right) - \left( \frac{\chi C_{\upsilon}}{C_p}\right) \frac{\mathrm{d} p_0}{\mathrm{d} t},
\end{equation}
where $C_{\upsilon}$ is the heat capacity at constant volume. Equation~\eqref{eq:divU_simplified} shows that the local volumetric expansion or contraction of the fluid is governed by the local heat fluxes, represented by the first term on the right-hand side, and the temporal evolution of the thermodynamic pressure, represented by the second term.

The thermodynamic pressure evolution is obtained by integrating Eq.~\eqref{eq:divU_simplified} over the entire domain. For a closed system with impermeable boundaries, the net velocity flux through the boundary is zero, giving
\begin{equation}\label{eq:dP0dt}
    \frac{\mathrm{d} p_0}{\mathrm{d} t} = \cfrac{  \displaystyle\int_V \cfrac{\beta}{\rho C_p}\ \cfrac{\partial }{\partial x_i}\left(\lambda \cfrac{\partial T}{\partial x_i}\right)\ \mathrm{d}V + \int_V  \cfrac{\beta}{\rho C_p} \ \dot{\mathcal{Q}}\ \mathrm{d}V}  {\displaystyle\int_V \chi\ \cfrac{ C_{\upsilon}}{C_p}\  \mathrm{d}V},
\end{equation}
where $V$ is the total volume of the domain.

The preceding derivation highlights the central role of the thermodynamic pressure in the low-Mach-number formulation. Equations~(\ref{eq:mass}, \ref{eq:momentum}, \ref{eq:cons_T}) comprise five conservation laws which, together with the equation of state~(Eq.~\eqref{eq:eos}), yield a total of $6$ governing equations for the seven unknowns $\rho,\ u,\ v,\ w,\ T,\ p_0,\ \text{and}\ p_1$. By comparison with the classical incompressible flow formulation, it is evident that the low–Mach–number approximation introduces two additional variables (i.e., $\rho \ \text{and} \ p_0$).  Density is determined directly from the equation of state, and the zeroth-order momentum equation establishes spatial uniformity of $p_0$. However, an additional constraint is required to determine the thermodynamic pressure. 

Different constraints on the thermodynamic pressure are applicable for open and closed systems. For an open system, the thermodynamic pressure is constant in time and therefore is equal to the initial value
\begin{equation}\label{eq:p0_open}
    p_0(t) = p_0(0) = \text{constant}.
\end{equation}
On the other hand, the temporal variations of thermodynamic pressure become relevant in a closed system. Utilizing the constraint that the total mass of the closed system remains constant, the thermodynamic pressure in a closed system can be evaluated as the solution to the inverse problem
\begin{equation}\label{eq:tot_mass}
    M_0 = \int_{V} \rho(p_0(t), T(\mathbf{x}, t)) \ \mathrm{d}V = \text{constant}.
\end{equation}
In other words, the thermodynamic pressure is determined such that the total mass of the system $M_0$ remains constant in time. As we will see, applying this constraint in a numerical method for low-Mach-number flows of real gases is non-trivial and novel to the present work.

\section{Numerical method}\label{sec:num_me}

The framework is implemented within the incompressible finite-difference solver CaNS \citep{costa_fft-based_2018}, which has been extended to solve the low-Mach-number equations for non-ideal fluids (within a stable single-phase regime). The governing equations are discretized on a staggered Cartesian grid, with velocity components stored at face centers and scalar quantities at cell centers. Diffusive terms and the convective terms in the momentum equations are discretized using second-order central differences. The convective terms in the energy equation are instead discretized using a third-order WENO scheme to suppress dispersive oscillations near strong temperature gradients. Such oscillations were observed to compromise robustness and produce nonphysical local extrema in simulations with large temperature differences. The solver is parallelized using MPI and supports GPU-accelerated computing platforms.

The algorithm proceeds in two main stages. In the first stage, the thermodynamic state is advanced to the new time level by solving the temperature equation and subsequently evaluating the thermodynamic pressure. The updated state is then used to determine the remaining thermophysical properties, including density. In the second stage, the momentum equations are advanced using a projection method, with the velocity-divergence constraint enforced locally at the current time step.

Time integration is performed explicitly using Wray's low-storage third-order Runge--Kutta scheme. The governing equations are advanced in three Runge--Kutta stages, $k = 1,2,3$, presented below in semi-discrete form, where $k =1$ corresponds to a time level $n$ and $k = 3$ to $n+1$. Each time step begins with the solution of the temperature equation

\begin{subequations}\label{eq:temperature_rk}
\begin{equation}\label{eq:temp_adv}
    T^{k+1} = T^k + \Delta t \left(\alpha_k^{\mathrm{RK}} {\Psi}_T^k + \beta_k^{\mathrm{RK}}  {\Psi}_T^{k-1} \right),
\end{equation}
with
\begin{equation}\label{eq:temp_rhs}
    {\Psi}_T = \frac{1}{\rho C_p}\left(-\rho C_p\frac{\partial }{\partial x_i}(u_iT) + \rho C_p T \frac{\partial u_i}{\partial x_i} + \frac{\partial }{\partial x_i}\left(\lambda \frac{\partial T}{\partial x_i}\right) + \dot{\mathcal{Q}}  + \beta T\frac{\mathrm{d} p_0}{\mathrm{d} t}\right).
\end{equation}
\end{subequations}
At Runge--Kutta stage $k$, the value of $\mathrm{d}p_0/\mathrm{d}t$ entering ${\Psi}_T^k$ is evaluated explicitly from Eq.~\eqref{eq:dP0dt} using the thermodynamic state available at level $k$; the value stored from the preceding stage is used in ${\Psi}_T^{k-1}$. After $T^{k+1}$, $p_0^{k+1}$, and the associated properties have been updated, Eq.~\eqref{eq:dP0dt} is evaluated again to provide the divergence constraint at the current stage and the pressure-work source for the next stage.
The RK3 coefficients are given by $\alpha_k^{\mathrm{RK}} = \{8/15, 5/12, 3/4\}$ and $\beta_k^{\mathrm{RK}} = \{0,-17/60,-5/12\}$.

\subsection{Evaluation of thermodynamic pressure}
With the temperature field available at the new time level, the thermodynamic pressure ($p_0$) is determined from the appropriate thermodynamic constraint. For open systems, this is given by Eq.~\eqref{eq:p0_open}, whereas for closed systems it is given by Eq.~\eqref{eq:tot_mass}. For the latter, determining the thermodynamic pressure is equivalent to finding the root of a nonlinear residual such that the density field obtained from the equation of state satisfies the global mass-conservation constraint.

For ideal gases, density depends linearly on pressure, yielding an explicit expression for the thermodynamic pressure in terms of the total mass and temperature field \citep{nicoud_dns_1999, Demou2019AFlows}:
\begin{equation}\label{eq:ig_pre}
    p_0^{k+1} = \cfrac{M_0 R}{\displaystyle\int_V \cfrac{1} {T^{k+1}} \ \mathrm{d}V},
\end{equation}
where $T^{k+1}$ is the temperature at the current time step and $R$ is the specific gas constant.

For non-ideal fluids, however, the nonlinear equation of state necessitates the solution of a nonlinear equation. In the present work, the thermodynamic pressure is obtained using a Newton–Raphson method. Defining the residual
\begin{subequations}\label{eq:newton_pressure}
\begin{equation}\label{eq:NR_residual}
    \mathcal{F}(p_0) = M_0 - \int_V \rho(p_0,T^{k+1}) \ \mathrm{d}V,
\end{equation}
the Newton update is given by
\begin{equation}\label{eq:NR_F}
    p_0^{m+1} = p_0^{m} - \frac{\mathcal{F}(p_0^{m})}{\mathcal{F}'(p_0^{m})},
\end{equation}
where $m$ denotes the Newton–Raphson iteration index. The initial iterate, $p_0^0$, is taken as the thermodynamic pressure from the previous time step ($p_0^k$), and the converged solution corresponds to $p_0^{k+1}$. $\mathcal{F}^{\prime}$ denotes the derivative of $\mathcal{F}$ with respect to $p_0$, which can be written as follows after swapping integration with differentiation:
\begin{equation}
    \mathcal{F}'(p_0) = -\int_V \left.\frac{\partial \rho}{\partial p_0} \right|_{T}\ \mathrm{d}V = -\int_V \rho \chi \ \mathrm{d}V,
\end{equation}
where the final equality follows from the definition of the isothermal compressibility, $\chi = (1/\rho)\left.\partial \rho/\partial p_0\right|_T$.
Substituting this into Eq.~\eqref{eq:NR_F} gives the final Newton update
\begin{equation}\label{eq:final_NR}
    p_0^{m+1} = p_0^{m} + \frac{M_0 - \displaystyle\int_V \rho\ \mathrm{d}V}{\displaystyle\int_V \rho\chi\ \mathrm{d}V}.
\end{equation}
\end{subequations}

In the ideal-gas limit, the equation of state is linear in thermodynamic pressure, and the Newton--Raphson procedure therefore converges in one iteration. Equation~\eqref{eq:final_NR} then reduces exactly to Eq.~\eqref{eq:ig_pre}, demonstrating that the proposed formulation recovers the classical ideal-gas result. For non-ideal fluids, the thermodynamic pressure from the previous time step provides a good initial guess. As a result, the Newton--Raphson procedure converges rapidly, requiring at most two to three iterations for all cases considered.

A key advantage of the proposed approach is that the derivative of the residual can be expressed in terms of isothermal compressibility, a quantity readily available from thermodynamic lookup tables. Consequently, the method is independent of the analytical form of the equation of state and can be coupled directly with tabulated thermodynamic data. This enables the framework to accommodate arbitrary real fluids within a stable single-phase regime, including those characterized by highly nonlinear constitutive relations of transport properties and caloric equation of state.

\subsection{Evaluation of thermophysical properties}
With the temperature and thermodynamic pressure available at the new time level, ($T^{k+1},\ p_0^{k+1}$), the thermodynamic state of the system is fully determined. The density is then obtained directly from the equation of state (Eq.~\eqref{eq:eos}), ensuring it is enforced exactly. The remaining thermophysical properties, including the thermal conductivity, specific heat capacities, coefficient of thermal expansion, and isothermal compressibility, are subsequently evaluated using the corresponding constitutive relations.

At each Runge--Kutta stage, changes in density give rise to a non-zero velocity divergence. Using the thermophysical properties evaluated at the current time level, the divergence constraint is computed locally within each control volume from Eqs.~\eqref{eq:divU_simplified} and \eqref{eq:dP0dt}. The resulting divergence field is then enforced through a projection step to obtain the updated velocity field. The thermodynamic state is thereby updated consistently, while conserving the total mass within the closed system. 

\subsection{Projection method}
With the density field available at the current time level, the momentum equations (Eq.~\eqref{eq:momentum}) are advanced using the incremental pressure--projection method of \citet{amsden1970simplified}, following the pressure-splitting approach of \citet{frantzis2019efficient} for variable-coefficient Poisson equations. Since the present implementation differs slightly from the standard formulation, the algorithm is described here for completeness. In this formulation, the hydrodynamic pressure is defined at half-time levels. A provisional velocity field is first obtained by integrating the momentum equations using the pressure from the previous half step.

The provisional velocity $u_i^*$ is computed as
\begin{subequations}\label{eq:provisional_velocity_update}
\begin{equation}\label{eq:pred_mom}
    u_i^{*} = \frac{\rho^k}{\rho^{k+1}} u_i^k - \gamma_k^{\mathrm{RK}} \Delta t\ \frac{1}{\rho^{k+1}} \frac{\partial p_1^{k-1/2}}{\partial x_i} +  \frac{\Delta t}{\rho^{k+1}}\left(\alpha_k^{\mathrm{RK}} \Psi _i^k + \beta_k^{\mathrm{RK}}  \Psi_i^{k-1} \right),
\end{equation}
where
\begin{equation}\label{eq:momentum_rhs}
    \Psi_i = - \frac{\partial }{\partial x_j}({\rho u_j u_i})  + \frac{\partial \tau_{ij}}{\partial x_j} + \rho f_i,
\end{equation}
\end{subequations}
and the RK coefficient $\gamma_k^{\mathrm{RK}} = \alpha_k^{\mathrm{RK}} + \beta_k^{\mathrm{RK}}$.

The pressure gradient term in the momentum equations is multiplied by the reciprocal of the density, which varies spatially. Retaining this form would result in a variable-coefficient Poisson equation. However, the computational efficiency of CaNS \citep{costa_fft-based_2018} relies on FFT-based solution of constant-coefficient Poisson equations. Hence, alternative formulations are considered to recast the problem into a constant-coefficient Poisson equation.

One possibility is to reformulate the projection step in terms of the momentum, $\rho^{k+1}u_i^*$, which yields the pair
\begin{subequations}\label{eq:conservative_projection}
\begin{equation}\label{eq:conservative_pressure_equation}
    \frac{\partial}{\partial x_i}\left(\rho^{k+1}u_i^*\right) - \frac{\partial}{\partial x_i}\left(\rho^{k+1}u_i^{k+1}\right) = \gamma_k^{\mathrm{RK}}\Delta t\,\frac{\partial^2 }{\partial x_i^2}\left(p_1^{k+1/2} - p_1^{k-1/2}\right),
\end{equation}
with
\begin{equation}\label{eq:conservative_mass_constraint}
    \frac{\partial}{\partial x_i}\left(\rho^{k+1}u_i^{k+1}\right) = -\frac{\partial \rho}{\partial t}.
\end{equation}
\end{subequations}
This approach has the advantage of enforcing mass conservation in its conservative form, with ${\partial \rho}/{\partial t}$ approximated using backward difference schemes \citep{mcmurtry_direct_1986,Cook1996DirectComputer,Najm1998AChemistry}. However, \citet{nicoud_conservative_2000} demonstrated that the method does not recover the divergence-free constraint in the inviscid limit, leading to a loss of kinetic-energy conservation. Numerical instabilities were also observed for temperature ratios exceeding approximately $3$. This limitation is particularly relevant to the present work, which focuses on flows in which large temperature variations give rise to significant density variations and strong gradients in thermophysical properties (e.g., non-ideal fluids), motivating the use of an alternative formulation,

\begin{subequations}\label{eq:pressure_splitting_scheme}
\begin{equation}\label{eq:pressplit}
    \frac{1}{\rho^{k+1}} \frac{\partial p_1^{k-1/2}}{\partial x_i}\rightarrow \frac{1}{\rho_*}\frac{\partial p_1^{k-1/2}}{\partial x_i} + \left(\frac{1}{\rho^{k+1}} - \frac{1}{\rho_*}\right)\frac{\partial p_1^*}{\partial x_i},
\end{equation}
where $\rho_*$ is a reference density\footnote{This value cannot be arbitrary: for numerical stability it should be equal to or lower than the minimum density in the domain \citep{Dong2012ARatios,Demou2019AFlows}.}, and \(p_1^*\) is an estimated hydrodynamic pressure calculated through linear extrapolation from the previous two time steps:
\begin{equation}\label{eq:pressure_extrapolation}
    p_1^* = \left(1 + \sum_{l=1}^k \gamma_l^{\mathrm{RK}} \frac{\Delta t^n}{\Delta t^{n-1}}\right)\ p_1^{n-1/2} - \sum_{l=1}^k \gamma_l^{\mathrm{RK}}\frac{\Delta t^n}{\Delta t^{n-1}}\ p_1^{n-3/2}.
\end{equation}
\end{subequations}
Here, the superscript $n$ corresponds to the time step and $k$ indicates the RK sub-step. When employing the pressure–splitting scheme within an RK time integration, care must be taken to ensure that the pressure is extrapolated consistently from previous time levels to the start of each substep. 

A similar pressure-splitting strategy was also employed by \citet{frantzis2019efficient} for incompressible two-fluid flows and by \citet{Motheau2016AAccuracy} for low-Mach-number reactive flows.The velocity correction, correction-pressure definition, and corresponding Poisson equation are
\begin{subequations}\label{eq:correction_system}
\begin{equation}\label{eq:velocity_correction_relation}
    \frac{ u_i^{k+1} - u_i^{*}}{\Delta t} = -\frac{\gamma_k^{\mathrm{RK}}}{\rho_*}\frac{\partial \varphi}{\partial x_i};
\end{equation}
\begin{equation}\label{eq:correction_pressure}
    \varphi =  p_1^{k+1/2} -  p_1^{k-1/2};
\end{equation}
\begin{equation}\label{eq:pressure_poisson}
    \cfrac{\partial^2 \varphi}{\partial x_i^2} = -\frac{\rho_*}{\gamma_k^{\mathrm{RK}} \Delta t} \left(\cfrac{\partial u_i^{k+1}}{\partial x_i} - \cfrac{\partial u_i^{*}}{\partial x_i}\right),
\end{equation}
\end{subequations}
where $\varphi$ is the correction pressure and Eq.~\eqref{eq:divU_simplified} prescribes the divergence of velocity locally at the current time step (${\partial u_i^{k+1}}/{\partial x_i}$). Hence, the pressure correction is obtained from a constant-coefficient Poisson equation, which is solved efficiently using FFT-based methods \citep{schumann1988fast}.

Finally, the velocity and pressure fields are corrected as
\begin{subequations}\label{eq:projection_corrections}
\begin{equation}\label{eq:velocity_correction}
    u_i^{k+1} = u_i^* -\frac{\gamma_k^{\mathrm{RK}}\Delta t}{\rho_*}\frac{\partial \varphi}{\partial x_i},
\end{equation}
and
\begin{equation}\label{eq:pressure_correction}
    p_1^{k+1/2} = p_1^{k-1/2} + \varphi.
\end{equation}
\end{subequations}

\subsection{Summary of the proposed method}

The sequence of steps in the algorithm can be summarized as follows:
\begin{enumerate}
    \item \emph{Initialization.}
    The velocity ($u_i$) and temperature ($T$) fields are initialized, and the initial thermodynamic pressure ($p_0$) is prescribed. Given $T$ and $p_0$, the density ($\rho$) is evaluated from the equation of state (Eq.~\eqref{eq:eos}), and the remaining thermophysical properties are obtained from the corresponding constitutive relations. For closed systems, the total mass of the system ($M_0$) is computed using Eq.~\eqref{eq:tot_mass}. The source term $\mathrm{d}p_0/\mathrm{d}t$ (Eq.~\eqref{eq:dP0dt}) and the velocity divergence field (Eq.~\eqref{eq:divU_simplified}) are evaluated using the initial thermodynamic state.
    \item \emph{Advance the temperature field.}  
    The temperature equation is advanced from $T^k$ to $T^{k+1}$ using Eq.~\eqref{eq:temp_adv}.
    \item \emph{Update the thermodynamic pressure.}  
    The thermodynamic pressure at the new substep, $p_0^{k+1}$, is computed using the Newton–Raphson iteration defined in Eq.~\eqref{eq:NR_F}.
    \item \emph{Update density and other thermophysical properties.}  
    Using the updated temperature field and thermodynamic pressure, the density and all other thermophysical properties are updated through the equation of state and the relevant constitutive relations.
    \item \emph{Compute the velocity divergence.}
    The ordinary time derivative of the spatially uniform thermodynamic pressure and the local velocity divergence are evaluated using Eqs.~\eqref{eq:dP0dt} and \eqref{eq:divU_simplified}, respectively. thermophysical properties at the current time step are used, yielding the divergence field for the current sub-step.
    \item \emph{Advance the velocity field.}  
    The momentum equations are advanced, and the velocity field $u_i^{k+1}$ is obtained using the pressure-projection method. 
\end{enumerate}

The above sequence is repeated at each RK sub-step until the target final time or statistical stationarity is reached. In what follows, we present a verification (\S~\ref{sec:MMS}) and validation (\S~\ref{sec:results}) of the computational solver. %

\section{Verification using manufactured solutions}\label{sec:MMS}

The method of manufactured solutions \citep{knupp2002verification, ShunnAN2007MethodSolvers} is used to verify the correctness of the numerical method and solver, including whether the observed order of accuracy is consistent with that expected from the discretization. %
This approach constructs analytical fields that satisfy the boundary conditions, which are substituted into the governing equations to generate residual source terms. These source terms are incorporated into the solver, and the resulting numerical solution is compared against the manufactured fields. Repeating the procedure on successively refined grids enables measuring the formal order of accuracy of the method.

In the present work, the method of manufactured solutions is employed to verify the coupling between the evolution of the thermodynamic state and the fluid motion. Particular emphasis is placed on the dynamic evolution of the thermodynamic pressure, the implementation of nonlinear equations of state, and the low–Mach–number projection algorithm that couples the thermodynamic and hydrodynamic fields through the velocity-divergence constraint.

\subsection{The manufactured solutions}

The solver is verified for the Van der Waals (VdW) equation of state (EoS) (Eq.~\eqref{eq:VdWeos}) using the method of manufactured solutions. The manufactured fields are constructed such that the velocity divergence is consistently related to the temporal variation of thermodynamic pressure and local heat fluxes, as described by Eq.~\eqref{eq:divU_simplified}. This approach is consistent with \citet{ShunnAN2007MethodSolvers}, who also construct manufactured solutions that satisfy mass conservation.

Since the governing equations include a transport equation for temperature, one might naturally seek to prescribe a manufactured temperature field directly (for instance, \citet{accary20063d}). For ideal gases, this is straightforward because the equation of state is linear in pressure for a given temperature, allowing the corresponding density field to be obtained analytically. For real fluids, however, a key challenge in constructing manufactured solutions is ensuring thermodynamic consistency between the density, pressure, and temperature fields. The nonlinear equation of state generally precludes a simple analytical construction of the density field from a prescribed temperature field, making it more involved to derive a consistent manufactured solution. 

Instead, for cubic equations of state such as the VdW EoS, it is more convenient to prescribe the density and thermodynamic-pressure fields and recover the temperature from the equation of state. Accordingly, the thermodynamic state is constructed from prescribed density and thermodynamic-pressure fields.

The manufactured solution is defined over a two-dimensional domain of size $L_x \times L_y$ with periodic boundary conditions in both directions, representing a closed system. The prescribed fields are therefore chosen to satisfy these boundary conditions. The manufactured density field is prescribed as
\begin{equation}\label{eq:manu_density}
    \rho_M(x,y,t) = \widetilde{\rho}_M \left(1 + A_{\rho} e^{-\sigma_{\rho} t} \cos(k_xx)\cos(k_y y)\right).
\end{equation}
Here, $A_{\rho}$ is the amplitude, $\sigma_{\rho}$ is the temporal decay rate, and $k_x = 2\pi m_x/L_x$ and $k_y = 2\pi m_y/L_y$ are the wavenumbers in the $x$- and $y$-directions, respectively, with $m_x$ and $m_y$ denoting the corresponding Fourier mode numbers. It can be easily verified that this solution has constant mass in this closed domain, as the spatial average of this solution is steady and equal to $\widetilde{\rho}_M$. 

The prescribed thermodynamic pressure needs to be a function of time only and is given by
\begin{equation}\label{eq:manu_pressure}
    p_{0,M}(t)=\widetilde{p}_{0,M}\left(1 + A_p e^{-\sigma_p t}\right),
\end{equation}
where $\widetilde{p}_{0,M}$ is the reference thermodynamic pressure, $A_p$ is the amplitude of the perturbation, and $\sigma_p$ is the rate at which the thermodynamic pressure drops. 

Equations~\eqref{eq:manu_density} and \eqref{eq:manu_pressure} are then used with the VdW EoS to evaluate the manufactured temperature field. In reduced variables, the VdW EoS is
\begin{equation}\label{eq:VdWeos}
    p_{\mathrm{r}} = \frac{8 \rho_{\mathrm{r}} T_{\mathrm{r}}}{3 - \rho_{\mathrm{r}}} - 3\rho_{\mathrm{r}}^2,
\end{equation}
where $p_{\mathrm{r}} = p/p_{\mathrm{cr}}$, $T_{\mathrm{r}} = T/T_{\mathrm{cr}}$, and $\rho_{\mathrm{r}} = \rho/\rho_{\mathrm{cr}}$ denote reduced pressure, temperature, and density, respectively, and the subscript `${\mathrm{cr}}$' corresponds to their value at the critical point. The manufactured fields in Eqs.~\eqref{eq:manu_density} and \eqref{eq:manu_pressure} correspond to reduced density and pressure fields. Accordingly, the manufactured temperature field is obtained using Eq.~\eqref{eq:VdWeos}. These prescribed thermodynamic fields ensure consistency with the equation of state and the thermodynamic constraints of the system.

The manufactured velocity field is consistently constructed by expressing the mass flux as the gradient of a scalar field. The resulting construction is
\begin{subequations}\label{eq:manu_velocity}
\begin{equation}\label{eq:manu_mass_flux}
    \rho_M \mathbf{u}_M = \nabla \Phi,
\end{equation}
which transforms the continuity equation into the Poisson equation
\begin{equation}\label{eq:manu_potential_poisson}
    \nabla^2 \Phi = -\frac{\partial \rho_M}{\partial t}.
\end{equation}
For the prescribed density field, the analytical expression for $\Phi$ is
\begin{equation}\label{eq:manu_potential}
    \Phi(x,y,t) = \frac{\dot{\rho}_M}{k^2},
\end{equation}
where $k^2 = k_x^2 + k_y^2$ and $\dot{\rho}_M = \partial \rho_M/\partial t$. An arbitrary divergence-free component may additionally be introduced through a streamfunction
\begin{equation}\label{eq:manu_streamfunction}
    \Psi_M = A_{\psi}e^{-\sigma_{\psi} t}\sin(k_x x)\sin(k_y y),
\end{equation}
where $A_{\psi}$ is the amplitude and $\sigma_{\psi}$ is the decay rate.
The resulting manufactured velocity field is therefore
\begin{equation}\label{eq:manu_mass_flux_complete}
    \rho_M\mathbf{u}_M = \nabla\Phi +
    \begin{bmatrix} 
    {\ \ \partial_y\Psi_M}\\
    -\partial_x\Psi_M
\end{bmatrix},
    \qquad
    \mathbf{u}_M = (u_M, v_M)^{\mathrm T}.
\end{equation}
Substituting the expressions for $\Phi$ and $\Psi_M$ yields
\begin{align}
    u_M(x,y,t) &= \frac{U(t)\sin(k_xx)\cos(k_yy)}{\rho_M},
    & U(t) &= \widetilde{\rho}_M A_{\rho}\sigma_{\rho}e^{-\sigma_{\rho}t}\frac{k_x}{k^2} + A_{\psi}e^{-\sigma_{\psi}t}k_y;
    \label{eq:manu_u}\\
    v_M(x,y,t) &= \frac{V(t)\cos(k_xx)\sin(k_yy)}{\rho_M},
    & V(t) &= \widetilde{\rho}_M A_{\rho}\sigma_{\rho}e^{-\sigma_{\rho}t}\frac{k_y}{k^2} - A_{\psi}e^{-\sigma_{\psi}t}k_x.
    \label{eq:manu_v}
\end{align}
\end{subequations}

The manufactured density, thermodynamic pressure, temperature, and velocity fields are now fully prescribed. The remaining properties, namely the viscosity, thermal conductivity, and heat capacity, are held constant. However, these fields do not satisfy Eqs.~\eqref{eq:momentum} and \eqref{eq:enthalpy} exactly. Substituting them into the governing equations therefore produces non-zero residuals, which are incorporated as the source terms in the momentum and temperature transport equations. Consequently, this verification exercises the nonlinear equation of state and the resulting variations of beta and chi, while the variable transport and caloric properties of the non-ideal-fluid cases are assessed through the validation studies of Sec.~\ref{sec:results}.

\subsection{Results}

The verification tests are performed on a domain of size $L_x \times L_y = 1\times 1$, initially discretized with $n_x \times n_y = 16\times 16$ control volumes, resulting in uniform grid spacing of $h = \Delta x = \Delta y$. The simulations are subsequently repeated on progressively refined uniform grids with $32$, $64$, $128$, and $256$ cells per direction, with $h$ decreasing accordingly. A time step of $\Delta t \propto h^2$ is employed across grids to ensure that temporal errors are insignificant relative to the spatial-discretization error. The parameters used to construct the manufactured solutions are summarized in Table~\ref{tab:parameters}. The thermophysical properties are held constant at $\mu = 1.68 \times 10^{-5}\, \mathrm{kg\,m^{-1}s^{-1}}$, $\lambda = 2.38 \times 10^{-2} \, \mathrm{W\,m^{-1}K^{-1}}$, and $C_p = 1004.5\, \mathrm{J\,kg^{-1}K^{-1}}$.

\begin{table}[!h]
\begin{center}
\setlength{\tabcolsep}{10pt}
\renewcommand{\arraystretch}{1.5}
\begin{tabular}{c c c | c c c | c c | c c}
\hline
$\widetilde{\rho}_M $ &
$A_{\rho} $ &
$\sigma_{\rho} $ &
$\widetilde{p}_{0,M} $ &
$A_{p} $ &
$\sigma_{p} $ &
$A_{\psi} $ &
$\sigma_{\psi} $ &
$m_x $ &
$m_y $ 
\\
\hline
\hline
$0.70$ &
$0.05$ &
$1.00$ &
$1.20$ &
$0.05$ &
$0.70$ &
$0.04$ &
$0.00$ &
$1.00$ &
$2.00$
\\
\hline
\end{tabular}
\caption{Parameters used to construct the manufactured solutions in the current study.}
\label{tab:parameters}
\end{center}
\end{table}

\begin{figure}
    \centering
    \includegraphics[width=\linewidth]{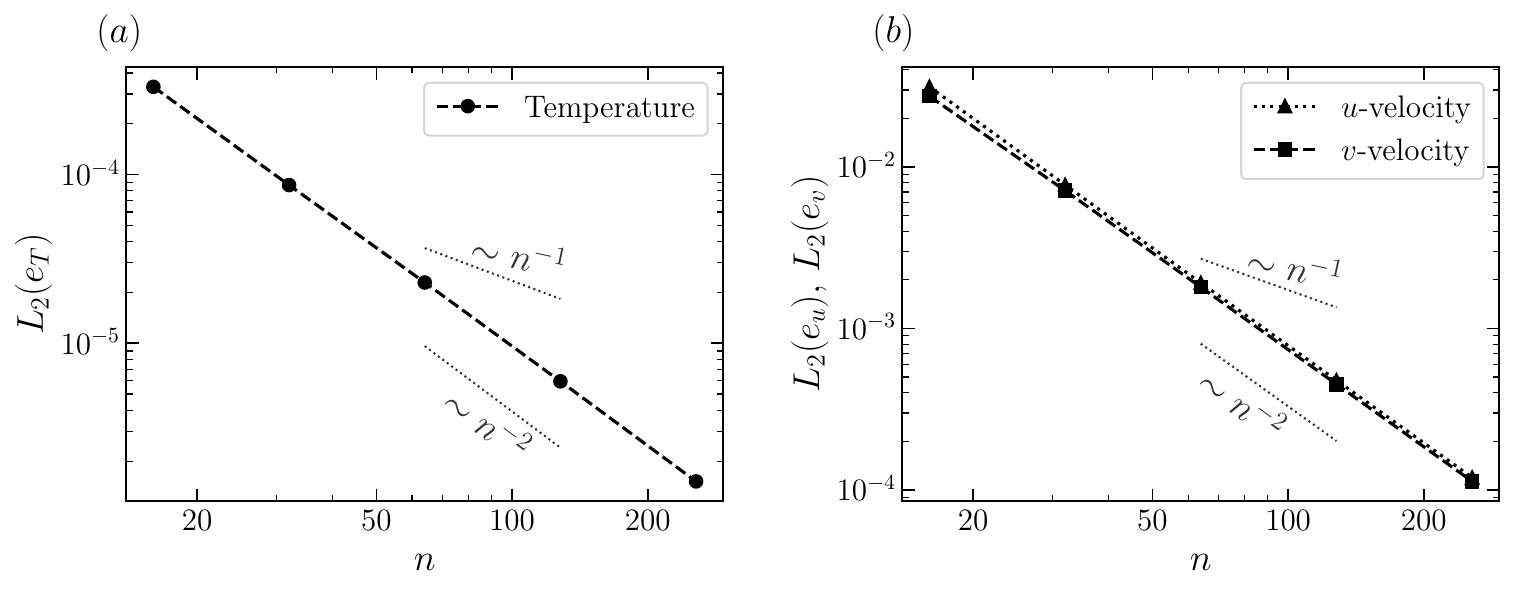}
    \caption{$L_2$-norm of the error for temperature and velocity components as a function of grid resolution at $t = 0.5$. The solver exhibits second-order spatial convergence.}
    \label{fig:MMS_OOC_Tuw}
\end{figure}

Figure~\ref{fig:MMS_OOC_Tuw} shows the $L_2$-norm of the error for temperature $L_2(e_T)$ and velocity components $L_2(e_u),\, L_2(e_v)$ as a function of the grid resolution at $t = 0.5$. %
Table~\ref{tab:na_mms_spatial_convergence_combined} summarizes the corresponding $L_2(e)$ error values, evaluated at $t = 0.5$. The solver exhibits the expected second-order spatial convergence for both velocity and temperature, thereby verifying the correctness of the implementation.

\begin{table}[!h]
\begin{center}
\setlength{\tabcolsep}{3.5pt}
\renewcommand{\arraystretch}{1.5}
\begin{tabular}{@{}cccccccc@{}}
\toprule
$n$ &
$\Delta t \times 10^{-4}$ &
$L_2(e_T) \times 10^{-5}$ &
$p_T$ &
$L_2(e_u) \times 10^{-3}$ &
$p_u$ &
$L_2(e_v) \times 10^{-3}$ &
$p_v$ \\
\hline
\hline
$16 $ & $39.062$ & $33.092$ & -- & $31.339$ & -- & $27.415$ & -- \\
$32 $ & $9.766$ & $8.657$ & 1.934 & $7.745$ & 2.017 & $7.138$ & 1.941 \\
$64 $ & $2.441$ & $2.290$ & 1.919 & $1.916$ & 2.015 & $1.802$ & 1.986 \\
$128 $ & $0.610$ & $0.594$ & 1.946 & $0.478$ & 2.003 & $0.452$ & 1.996 \\
$256 $ & $0.153$ & $0.152$ & 1.970 & $0.119$ & 2.000 & $0.113$ & 1.999 \\
\bottomrule
\end{tabular}
\caption{$L_2$-norm of the error and observed spatial convergence orders for temperature ($T$) and velocity components ($u, v$) for the manufactured case with the VdW EoS. The domain is discretized using $n\times n$ cells with the time step scaled as $\Delta t \sim h^2$, where $h$ is the grid spacing. The parameters $p_T$, $p_u$, and $p_v$ denote the computed spatial convergence orders ($\mathcal{O}(h^p)$). All errors are evaluated at $t=0.5$.}
\label{tab:na_mms_spatial_convergence_combined}
\end{center}
\end{table}

\section{Validation and applications of the proposed framework}\label{sec:results}
The low–Mach–number Navier--Stokes framework is validated against reference benchmarks with increasing complexity, from the laminar natural convection of an ideal gas, to the turbulent channel transport of supercritical $\mathrm{CO}_2$ near the critical point. %

\subsection{Differentially heated cavity}

The differentially heated cavity (DHC) is a two–dimensional configuration in which the two vertical walls are maintained at different constant temperatures, while the top and bottom walls are adiabatic \citep{le_quere_accurate_1991}. Gravity acts in the vertical direction. The flow is characterized by Rayleigh number $Ra = g \beta L^3 \Delta T/(\nu \alpha)$, $\beta = 1/T_0$ is the thermal expansion coefficient, $\nu$ is the kinematic viscosity, $\alpha$ is the thermal diffusivity, and $\Delta T$ is the prescribed temperature difference between the walls.

\subsubsection{Ideal gas}

The fluid density is related to temperature through the ideal gas (IG) law, and a natural convection loop develops within the cavity. Since the system is closed, the thermodynamic pressure evolves in time to satisfy the global mass conservation constraint. The cases feature high temperature differences that render the Oberbeck--Boussinesq approximation inapplicable.

The fluid is uniformly initialized at a reference temperature of $T_0 = 600\, \mathrm{K}$. The left wall is maintained at $T_1 = 960\, \mathrm{K}$, while the right wall is kept at $T_2 = 240\, \mathrm{K}$, resulting in a temperature difference of $\Delta T = 720\, \mathrm{K}$. The gravitational acceleration $g$ is prescribed such that the desired Rayleigh number $(Ra)$.
The fluid is a thermally and calorically perfect gas. The viscosity and thermal conductivity are given using Sutherland’s law, following \citet{Demou2019AFlows}.

\begin{figure}
    \centering
    \includegraphics[width=\linewidth]{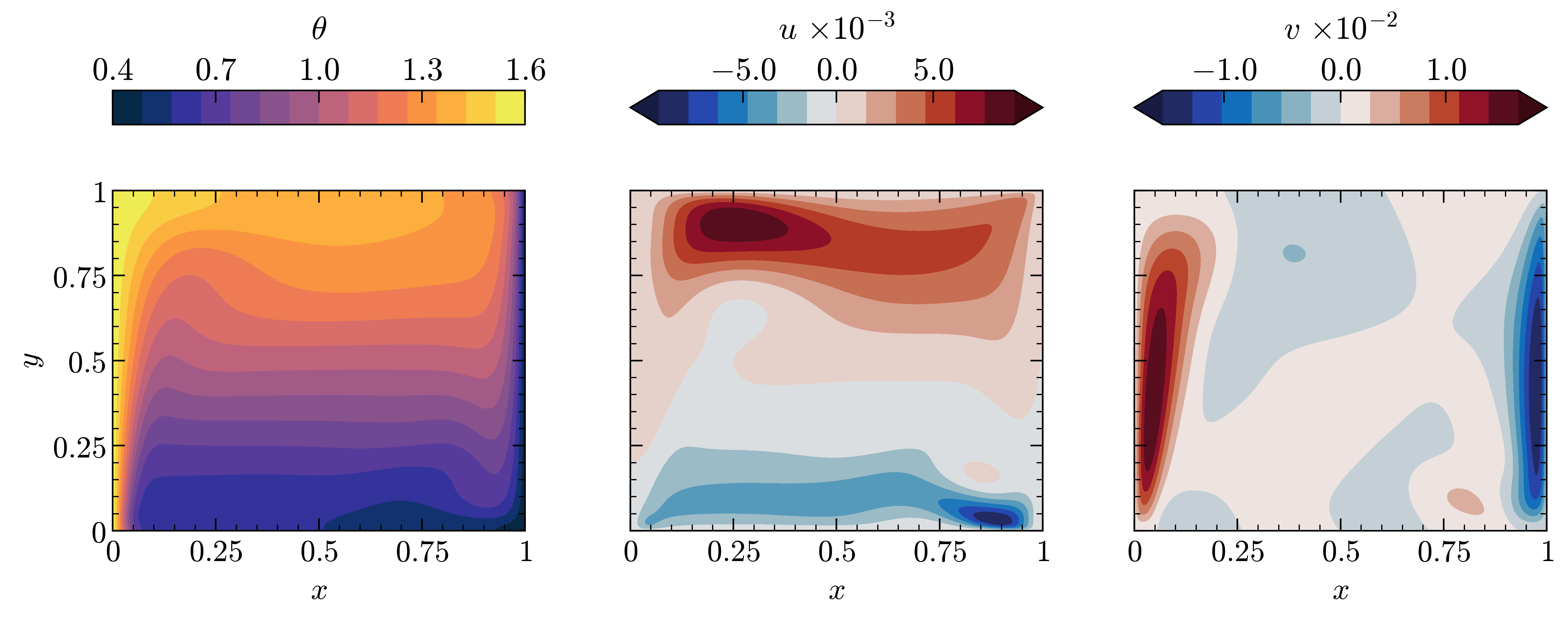}
    \caption{Contours of the normalized temperature field, $\theta = T/T_0$ (with $T_0 = 600\, \mathrm{K}$), horizontal velocity $u$, and vertical velocity $v$, from left to right, for a 2D differentially heated cavity with ideal gas at $Ra = 10^6$. The vertical walls are maintained at a temperature difference of $\Delta T = 1.2\, T_0$.}
    \label{fig:2d_DHC_contours}
\end{figure}

\begin{figure}
    \centering
    \includegraphics[width=\linewidth]{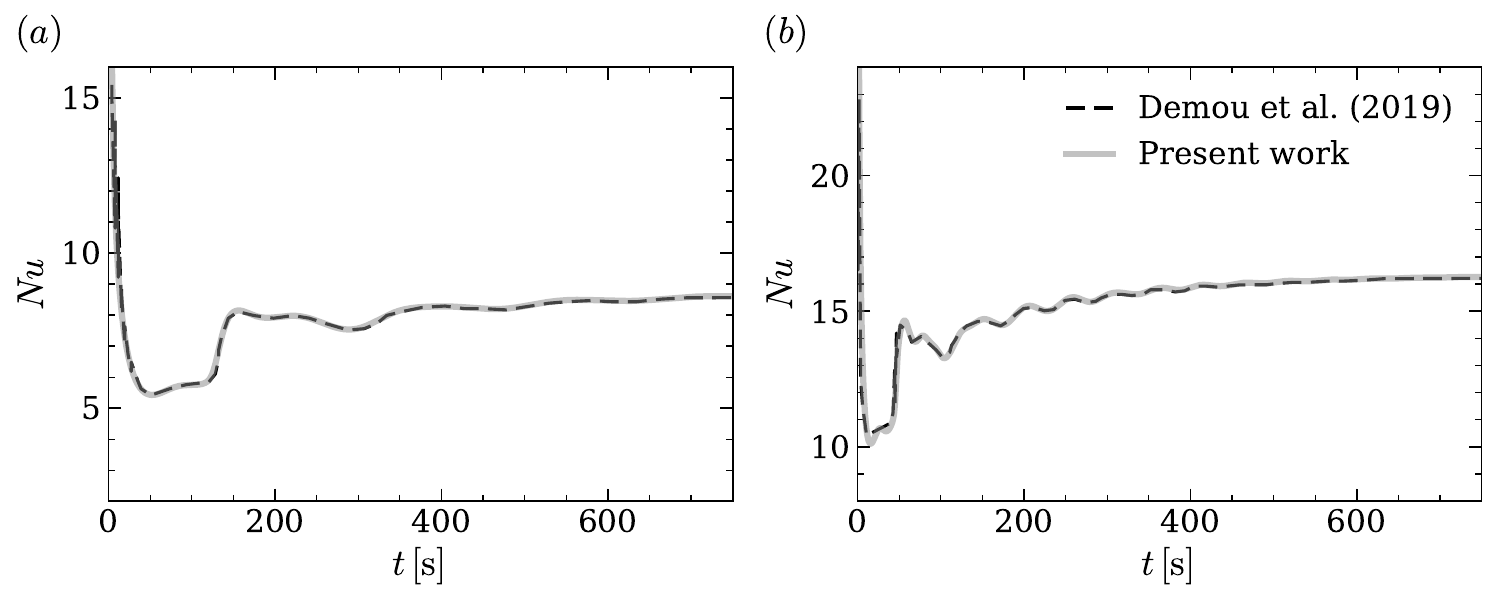}
    \caption{Temporal evolution of the Nusselt number for a 2D differentially heated cavity with ideal gas at (\emph{a}) $Ra = 10^6$ and (\emph{b}) $Ra = 10^7$. The temperature difference between the walls is $\Delta T = 720\, \mathrm{K}$. Present results (solid lines) are compared with the data of \citet{Demou2019AFlows} (dashed lines). }
    \label{fig:2d_DHC_evol_Nu}
\end{figure}

The simulations were carried out at $Ra = 10^6$ and $Ra = 10^7$ in a square cavity of dimensions $L_x \times L_y = 1\, \mathrm{m} \times 1\, \mathrm{m}$ and $Pr = 0.71$. Starting from a stationary state, the flow was evolved until a steady state was attained, on a sequence of progressively refined uniform grids to assess grid independence. Figure~\ref{fig:2d_DHC_contours} presents the steady-state temperature field together with the horizontal and vertical velocity components on a $256 \times 256$ grid. The differential heating induces the characteristic buoyancy-driven convective loop within the cavity, with rising fluid adjacent to the hot wall and descending fluid along the cold wall. Figure~\ref{fig:2d_DHC_evol_Nu} compares the temporal evolution of the Nusselt number ($Nu = q L / (\lambda \Delta T)$) in the present study with the results of \citet{Demou2019AFlows}, demonstrating that the solver accurately captures the transient dynamics. Additionally, Table~\ref{tab:DHC_2d} reports the steady-state thermodynamic pressure ($p_0$) and the Nusselt number ($Nu$) obtained on the different grids, alongside benchmarks from \citet{LeQuere2005ModellingSolutions}. The predicted Nusselt numbers converge monotonically with grid refinement and agree closely with the reference values, while the thermodynamic pressure also exhibits grid-independent behavior.

\begin{table}[!h]
\begin{center}
\setlength{\tabcolsep}{10pt}
\renewcommand{\arraystretch}{1.5}
\begin{tabular}{c | c | c | c | c | c | c | c }
\hline
& & \multicolumn{3}{c|}{$Nu$} & \multicolumn{3}{c}{$\widetilde p_0$}\\
\hline
 $Ra$                & Grid & Present & Reference & $\Delta (\%)$ & Present & Reference & $\Delta (\%)$\\
\hline
\hline
    \multirow{3}{*}{$10^6$}& $128\times 128$ & 8.7957   & \multirow{3}{*}{$8.6866$} & 1.256 & 0.92005   & \multirow{3}{*}{0.924487} & 0.480 \\
      & $256\times 256$ & 8.7173 &    & 0.354 & 0.92315 &    & 0.145 \\
      & $512\times 512$ & 8.6945 &    & 0.092 & 0.92414 &    & 0.037 \\
\hline
    \multirow{3}{*}{$10^7$} & $128\times 128$ & 16.7029   & \multirow{3}{*}{$16.2410$} & 2.844 & 0.91390   & \multirow{3}{*}{$0.92263$} & 0.946\\
      & $256\times 256$ & 16.4096 &    & 1.038 & 0.91831 &    & 0.468 \\
      & $512\times 512$ & 16.3253 &    & 0.519 & 0.92046 &    & 0.235 \\
\hline
\end{tabular}
\caption[Macroscopic quantities obtained from the simulations for different cases.]{\label{tab:DHC_2d}Comparison of steady-state values of thermodynamic pressure ($\widetilde p_0$) and the Nusselt number, $Nu = q_h L / (\lambda_h \Delta T)$, for a 2D differentially heated cavity with ideal gas. Here, the subscript $h$ denotes quantities evaluated at the hot wall, and $\widetilde p_0$ is the thermodynamic pressure normalized by its initial value. Reference values reported by \citet{LeQuere2005ModellingSolutions} are shown for the $\Delta T = 720\ \mathrm{K}$ case, with $T_h/T_c = 4$.}
\end{center}
\end{table}

For comparison, the DHC configuration is also simulated with a Van der Waals fluid using the property model described below.

\subsubsection{Van der Waals fluid}

The same DHC configuration is now simulated using the Van der Waals (VdW) equation of state, which accounts for both the finite molecular volume and intermolecular attractive forces. Expressed in terms of reduced variables, the thermal equation of state is given by Eq.~\eqref{eq:VdWeos}.

The system is initialized at a supercritical pressure ($p_0(0) = 1.35\,p_{\mathrm{cr}}$), and at the critical temperature, corresponding to a liquid-like state. The initial temperature is set to $T_0 = T_{\mathrm{cr}}$, and the left and right walls are maintained at $T_1 = T_{\mathrm{cr}} + \Delta T/2$ (gas-like state) and $T_2 = T_{\mathrm{cr}} - \Delta T/2$ (liquid-like state), respectively, where $\Delta T/T_{\mathrm{cr}} = 1.2$ is the same normalized relative temperature difference used in the IG case. The temperature difference yields a density ratio of $\rho_2/\rho_1 = 7.34$ between the cold and hot walls initially, which is higher than the density ratio of $4$ in the IG case. The larger wall-density ratio is expected to enhance buoyancy-driven motion compared with the IG case.

The dynamic viscosity and thermal conductivity are assumed to be constant and are evaluated for carbon dioxide at the initial pressure ($p_0(0) = 1.35~p_{\mathrm{cr}}$) and the critical temperature using the \emph{CoolProp} database \citep{coolprop2014}. The values used in the present simulation are $\mu = 6.49 \times 10^{-5}\ \mathrm{kg\,m^{-1}\,s^{-1}}$ and $\lambda = 8.28 \times 10^{-2}\ \mathrm{W\,m^{-1}\,K^{-1}}$.
The VdW thermal equation of state must be complemented by a thermodynamically consistent \emph{caloric} equation of state, here specified through the heat capacities. The heat capacities employed in the present simulations are described in Appendix~\ref{sec:cpvdw}. 

\begin{figure}
    \centering
    \includegraphics[width=\linewidth]{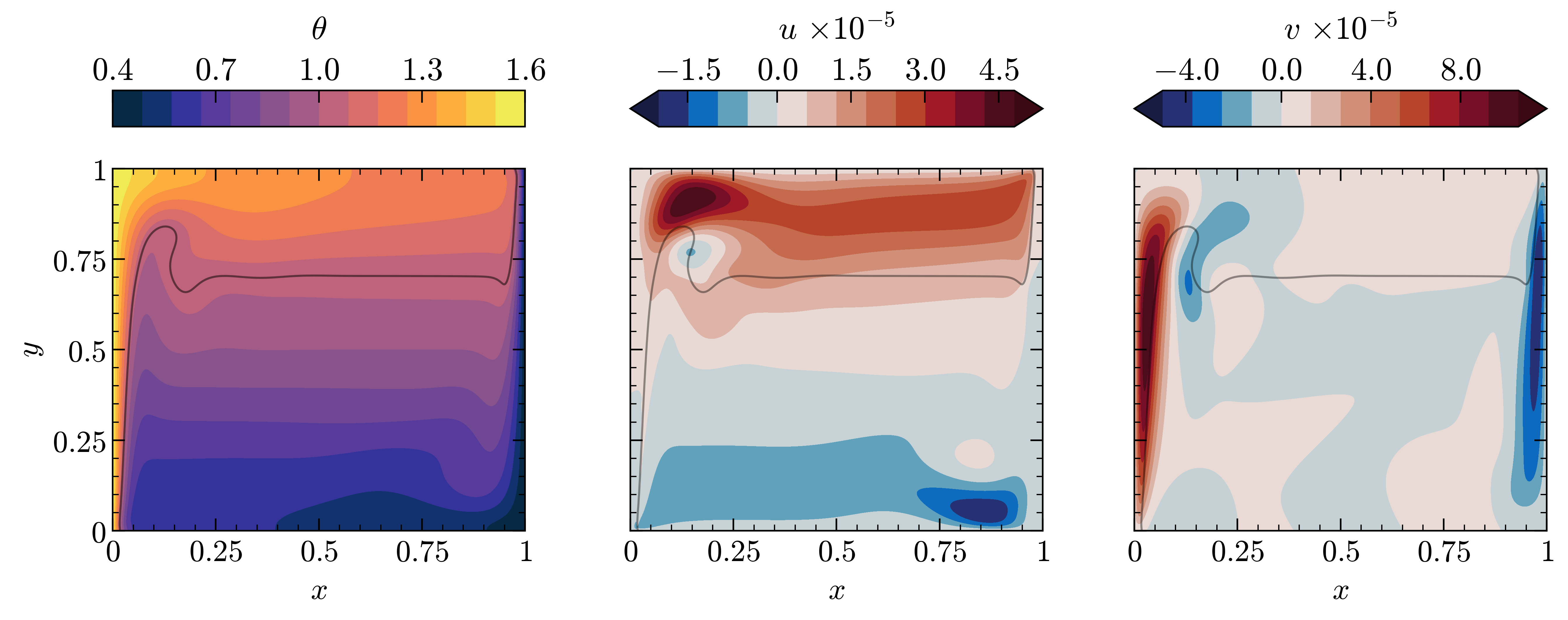}
    \caption{Contours of the normalized temperature field, $\theta = T/T_0$ (with $T_0 = T_{\mathrm{cr}}$), horizontal velocity $u$, and vertical velocity $v$, from left to right, for a 2D differentially heated cavity with the Van der Waals fluid at $Ra = 10^6$. The vertical walls are maintained at a temperature difference of $\Delta T = 1.2\, T_0$. The gray line in the contours indicates the location where the local thermodynamic state crosses the Widom line. Along this line, $\theta = 1.037$, corresponding to the pseudo-boiling temperature at $p_0 = 1.156\, p_{\mathrm{cr}}$.}
    \label{fig:DHC_VdW}
\end{figure}

The gravitational acceleration is prescribed to achieve the Rayleigh number ($Ra = 10^6$), matching that of the IG case, and the flow evolves under buoyancy until a steady state is attained. The cavity is simulated on a uniform grid of $512 \times 512$ points, and the steady-state results are compared against the IG case. Qualitatively, the temperature distribution in the cavity is similar to that of the IG case, as can be seen in the contours presented in Figure~\ref{fig:DHC_VdW}. The gray line in the temperature contour marks the locations where the local thermodynamic state crosses the Widom line, separating the liquid-like and gas-like regions of the supercritical fluid. The steady-state values of the Nusselt number and the thermodynamic pressure, however, differ from those of the IG case, as shown in Table~\ref{tab:DHC_vdW}. The Nusselt number, for the same normalized relative temperature difference ($\Delta T/T_0$), is higher for the VdW fluid, indicating a higher heat-transfer rate than in the IG case.  Figure~\ref{fig:DHC_VdW} also shows the horizontal and vertical velocity contours, which exhibit a pronounced asymmetry compared to the IG case. The gas-like region above the gray line exhibits higher velocities than the liquid-like region below it because of its lower density. These results demonstrate the influence of non-ideal behavior on the buoyancy-driven flow field.

\begin{figure}
    \centering
    \begin{overpic}[width=\linewidth]{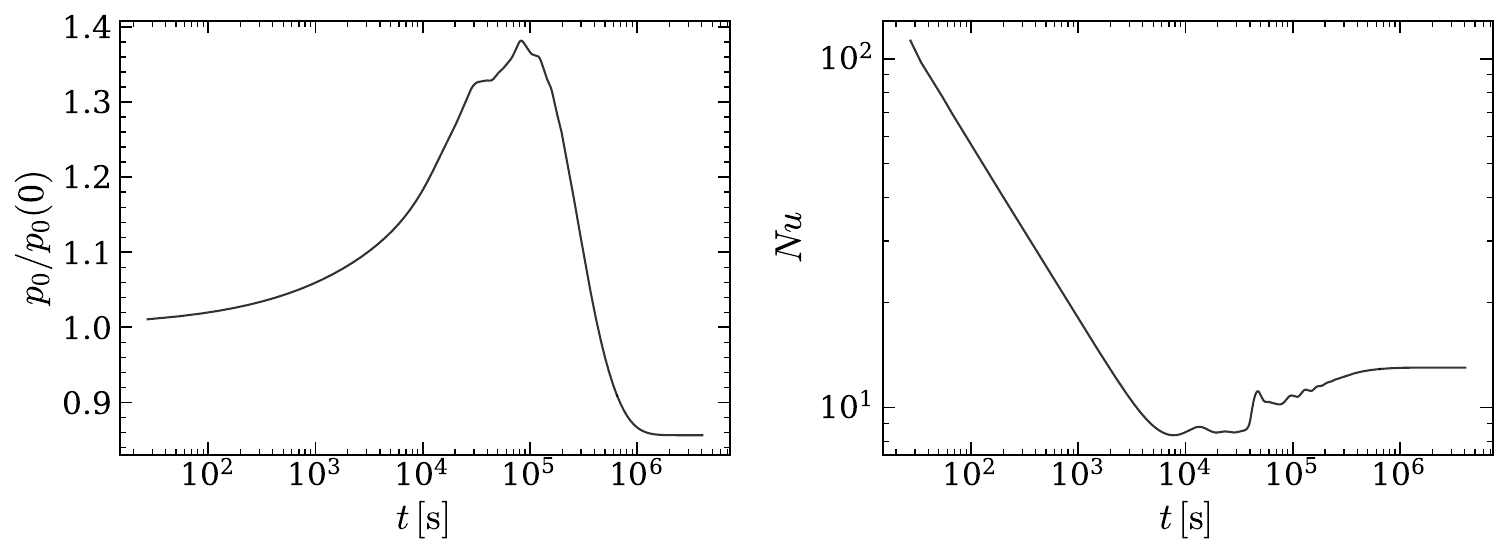}
        \put(0,35){(\emph{a})}
        \put(53,35){(\emph{b})}
    \end{overpic}
    \caption{Temporal evolution of (\emph{a}) the normalized thermodynamic pressure and (\emph{b}) the Nusselt number for a 2D differentially heated cavity with the Van der Waals fluid at $Ra = 10^6$. The temperature difference between the hot and cold walls is $\Delta T = 1.2\, T_{\mathrm{cr}}$.}
    \label{fig:DHC_VdW_transient}
\end{figure}

Additionally, the transient evolution of the Nusselt number and thermodynamic pressure is shown in Figure~\ref{fig:DHC_VdW_transient}. Compared with the IG case, the VdW fluid exhibits a much longer transient before reaching the steady state. Simulating this transient with a fully compressible solver would require a prohibitively small time step to resolve acoustic waves. For the present case, in fact, the characteristic Mach number is $\mathcal{O}(10^{-7})$, making such simulations with fully compressible solvers \citep{boldini2025cubens} computationally impractical.

This underscores the importance of the proposed low-Mach-number formulation for accurately predicting heat transfer in non-ideal fluids, particularly when non-ideal thermodynamic effects are significant, while avoiding the severe acoustic time-step restriction imposed by fully compressible methods.

\begin{table}[!h]
\begin{center}
\setlength{\tabcolsep}{10pt}
\renewcommand{\arraystretch}{1.5}
\begin{tabular}{c | c | c  }
\hline
              & IG & VdW  \\
\hline
\hline
$Nu$ & $8.6945$ & $13.0088$\\
\hline
$\widetilde p_0$ & 0.9241 & 0.8563\\
\hline
\end{tabular}
\caption[Macroscopic quantities obtained from the simulations for different cases.]{\label{tab:DHC_vdW}Comparison of steady-state values of thermodynamic pressure ($\widetilde p_0$) and the Nusselt number, $Nu = q_h L / (\lambda_h \Delta T)$, for an ideal gas and a VdW fluid. Here, the subscript $h$ denotes quantities evaluated at the hot wall, and $\widetilde p_0$ is the thermodynamic pressure normalized by its initial value. For the VdW fluid, the steady-state thermodynamic pressure corresponds to $1.156\, p_{\mathrm{cr}}$.}
\end{center}
\end{table}

\subsection{Turbulent channel flow}\label{sec:TCF}
The numerical framework developed in the present work is now validated against direct numerical simulations (DNS) of pressure-driven turbulent channel flow without gravity. The channel walls are maintained at prescribed constant temperatures, inducing substantial density variations across the flow, while periodic boundary conditions are imposed in the streamwise and spanwise directions. 

\subsubsection{Ideal gas}

For the ideal-gas case, the resulting mean-flow profiles are compared with the reference data of \citet{nicoud_conservative_2000}. Here, the bottom wall is held at a temperature $T_1$, while the top wall is maintained at $T_2$, such that $T_1 + T_2 = 2$. The non-dimensional viscosity and thermal conductivity are prescribed as $\mu = \lambda = \sqrt{p_0/T}$, where $p_0$ and $T$ denote the non-dimensional thermodynamic pressure and temperature, respectively, following \citet{nicoud_conservative_2000}. Simulations are performed for three temperature ratios, $T_2/T_1 = 1.01,\ 2,\ 4$.

\begin{figure}
    \centering
    \includegraphics[width=\linewidth]{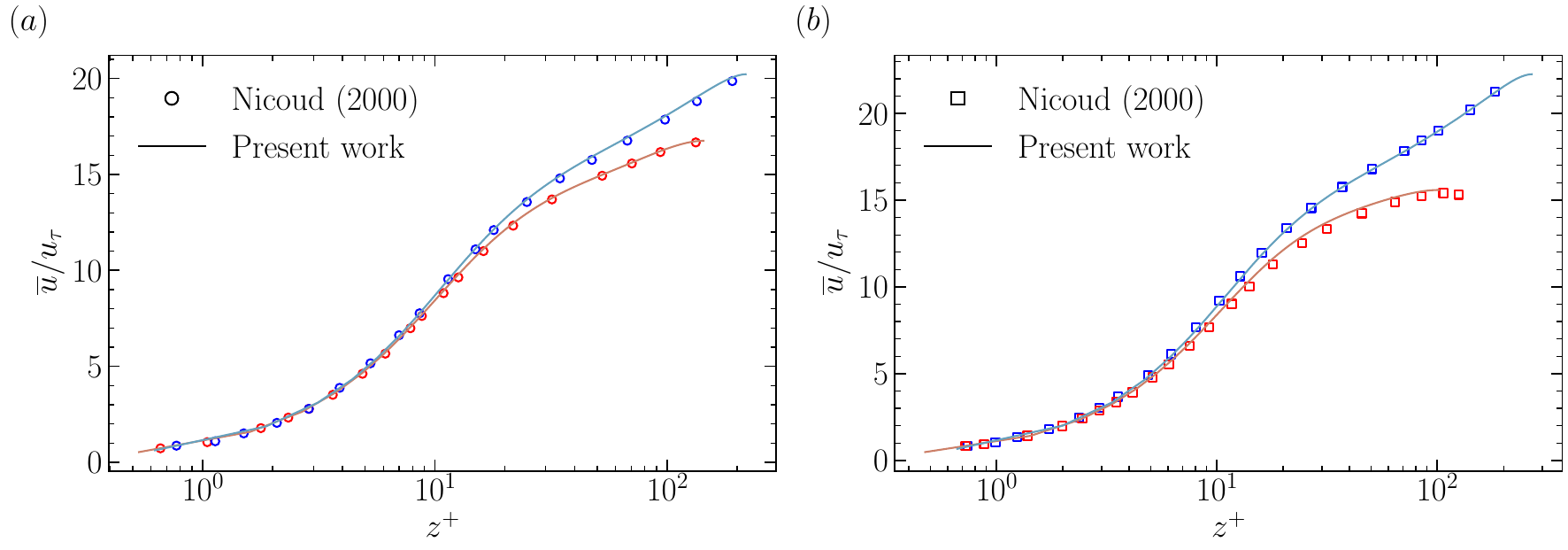}
    \caption{Mean-velocity profiles for ideal-gas turbulent channel flows at temperature ratios of (\emph{a}) $T_2/T_1  = 2$ and (\emph{b}) $T_2/T_1 = 4$, compared to the data of \citet{nicoud_conservative_2000}. Red and blue colors represent the hot and cold sides of the channel, respectively.}
    \label{fig:vel_TR2_TR4}
\end{figure}

The flow requires approximately 50 eddy turnover times $(h/u_{\tau})$ to reach a statistically steady state. Flow statistics are then averaged over no fewer than 150 samples collected during a period of 100 eddy turnover times. The mean-velocity profiles on the cold and hot sides of the channel are compared with the reference data in Figure~\ref {fig:vel_TR2_TR4}, showing very good agreement. In addition, the friction Reynolds numbers, skin-friction coefficients, and heat-flux parameters at both walls are summarized in Table~\ref{tab:valNic}. These integral quantities also agree closely with the reference values.

\begin{table}[!h]
\begin{center}
    \begin{tabular}{c|c| c| c| c| c| c| c| c }
    \hline
         &&&&&&&&\\
          &Case& \(Re_{\tau, c}\) & \(Re_{\tau, h}\) & \(Re_{b}\)& \(C_{f1} \times 10^3\) & \(C_{f2} \times 10^3\) & \(B_{q1}\)&\(B_{q2}\)\\
        &&&&&&&&\\
         \hline
         \hline
        &&&&&&&&\\
        Present code&& 180 & 180 & 2838 &6.03 & 5.98 & \(\approx 0\) & \(\approx 0\) \\
        &\(T_2/T_1 = 1.01\) &  &  &  &  &  &  &    \\
         \citet{nicoud_conservative_2000}&& 182 & 185 & 2855 & 6.1 & 6.1 & \(\approx 0\) & \(\approx 0\)  \\
        &&&&&&&&\\
        \hline
        &&&&&&&&\\
        Present code&& 194 & 165 & 2846 & 6.87 & 5.01 & -0.018 & 0.015  \\
         &\(T_2/T_1 = 2\)& &  &  &  &  &  &   \\
        \citet{nicoud_conservative_2000}&& 195& 164 & 2810 & 7.0 & 5.0 & -0.018 & 0.016     \\
        &&&&&&&&\\
        \hline
        &&&&&&&&\\
        Present code&& 206 & 148 & 2844 & 7.70 & 3.98 & -0.040 & 0.028 \\
        &\(T_2/T_1 = 4\) &  &  &  &  &  &  &    \\
         \citet{nicoud_conservative_2000}&& 211 & 151 & 2818 & 8.2 & 4.2 & -0.041 & 0.029  \\
        &&&&&&&&\\
        \hline
    \end{tabular}
\caption[Validation of macroscopic quantities obtained in the present code with \citet{nicoud_conservative_2000}.]{\label{tab:valNic}Validation of macroscopic quantities obtained in the present code with \citet{nicoud_conservative_2000}. The subscript \(1\) refers to the cold wall and \(2\) to the hot wall. \(C_f\) is the coefficient of friction based on mean density and maximum velocity. \(B_q\) is the wall heat flux parameter defined as \(B_q=q_w/\rho_w C_p u_{\tau}T_w\).}
\end{center}
\end{table}

\subsubsection{Supercritical $\mathrm{CO}_2$}

To further validate and demonstrate the capabilities of the present framework, a turbulent channel flow with a non-ideal fluid near its critical point is considered. In this regime, thermophysical properties exhibit strong nonlinearities and sharp gradients. Supercritical carbon dioxide ($\mathrm{CO}_2$) is selected as the working fluid, with critical pressure and temperature $p_{\mathrm{cr}} = 7.38\ \mathrm{MPa}$ and $T_{\mathrm{cr}} = 304.2\ \mathrm{K}$, respectively. The flow is initialized at a supercritical pressure of $p_0(t=0) = 8.4\ \mathrm{MPa}$, which yielded a steady-state thermodynamic pressure close to that reported in \cite{wan2025effects}. 

The lower and upper walls are maintained at $T_1 = 297.8\ \mathrm{K}$ and $T_2\ = 317.8\ \mathrm{K}$, respectively, corresponding to a temperature difference of $20\ \mathrm{K}$. These conditions are chosen to reproduce the configuration of \citet{wan2025effects}, who considered the same temperature difference at a constant supercritical pressure of $8\ \mathrm{MPa}$. In the present simulations, however, unlike \citet{wan2025effects}, the pressure is not constant and evolves during the course of the simulation. Hence, the simulation is set up such that the thermodynamic state once the flow is fully developed is close enough for meaningful comparison. Although fixing $p_0$ in the reference DNS \citep{wan2025effects} does not enforce the total-mass constraint of a strictly closed system, this approximation appears benign for the statistically stationary state considered here: the agreement shown below for both the mean profiles and the turbulent kinetic-energy budgets indicates that the associated mass discrepancy does not materially alter the reported turbulence statistics. The reference data therefore provide a useful and reassuring validation target, while the present formulation additionally guarantees global mass conservation throughout the transient and statistically stationary regimes.

The thermophysical properties at a given temperature and pressure are obtained from the \emph{CoolProp} library \citep{coolprop2014}, which provides accurate thermophysical properties for a wide range of fluids. The properties are input to the solver as tabulated data and interpolated at runtime. The temperature range of the tabulated data is $287.8 \mathrm{K} \leq T \leq 327.8\ \mathrm{K}$. The pressure range of the tabulated data is $7.45\ \mathrm{MPa} \leq p \leq 11.06\ \mathrm{MPa}$, which includes the initial pressure and the expected range of pressure variations during the simulation. The table is generated at a resolution of $0.04\ \mathrm{K}$ in temperature and $7.22\times 10^{-2}\ \mathrm{MPa}$ in pressure, resulting in $1000$ and $50$ data points for temperature and pressure, respectively. Following the work of \citet{rinaldi2014exact}, a third-order Lagrange interpolation is carried out to obtain the properties at intermediate thermodynamic states. This table and interpolation procedure are used for all transcritical simulations reported below.

\begin{figure}
    \centering
    \includegraphics[width=\linewidth]{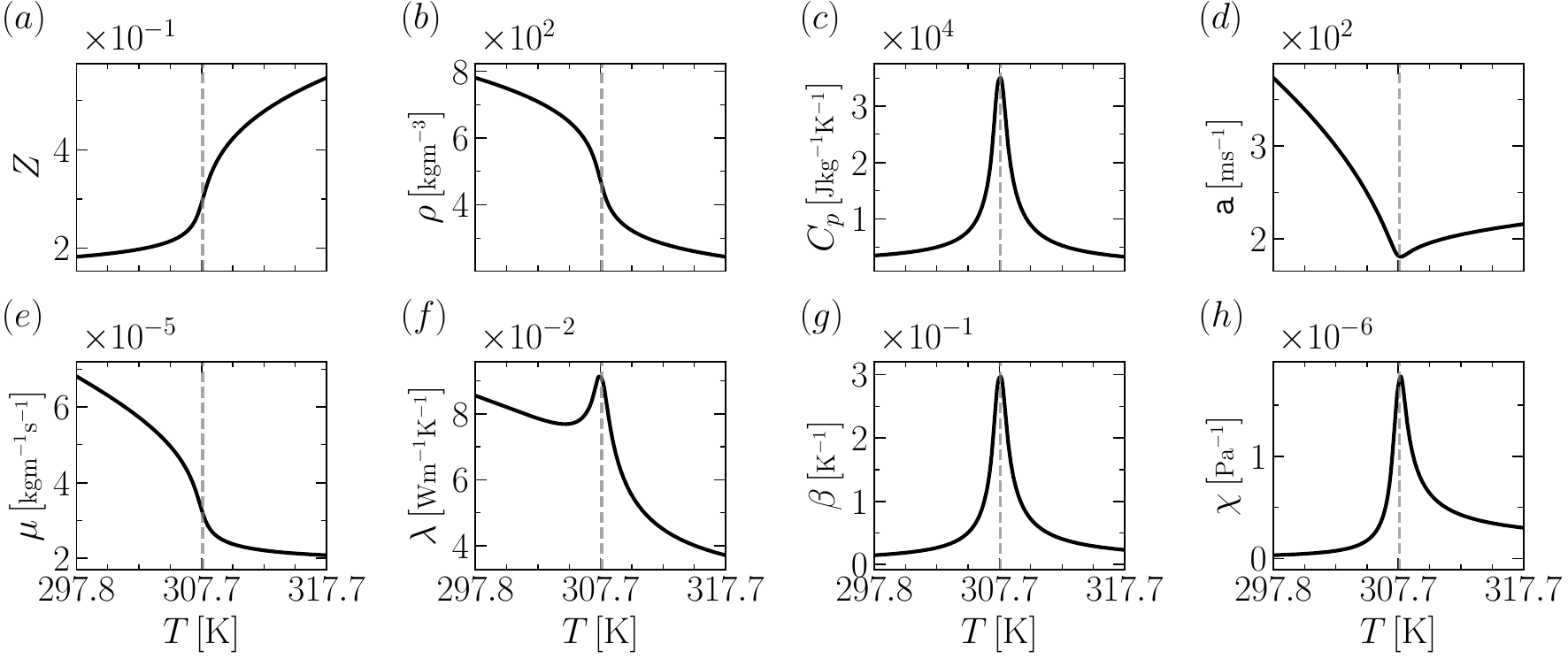}
    \caption{Thermophysical properties of $\mathrm{CO}_2$ as a function of temperature in the range $0.975 \leq T/T_{\mathrm{pb}} \leq 1.025$ at a constant pressure $p_{\mathrm{ref}} = 8\ \mathrm{MPa} \simeq 1.084\,p_{\mathrm{cr}}$: (\emph{a}) compressibility factor; (\emph{b}) density; (\emph{c}) specific heat capacity; (\emph{d}) speed of sound; (\emph{e}) viscosity; (\emph{f}) thermal conductivity; (\emph{g}) coefficient of thermal expansion; (\emph{h}) isothermal compressibility. The vertical dashed line denotes the pseudo-boiling temperature $T_\mathrm{pb}=307.7\ \mathrm{K}$.}
    \label{fig:prop_varsCO2_CP}
\end{figure}

The property distributions across $T_1$ and $T_2$ at $8\ \mathrm{MPa}$, corresponding to the fully-developed state, are illustrated in Figure~\ref{fig:prop_varsCO2_CP}. Strong thermodynamic nonlinearity is observed within the pseudo-boiling region, where the thermophysical properties vary over a narrow temperature range. In particular, both density and specific heat capacity vary sharply across the channel. The density ratio between the cold and hot walls is approximately $3$, while the ratio of the peak heat capacity to its wall value (cold side) approaches $10$. Similarly, the transport properties also exhibit steep variations as the fluid transitions from liquid-like to gas-like states. The compressibility factor increases from the cold wall to the hot wall, indicating that the fluid behavior becomes progressively more ideal across the channel.

\begin{figure}
    \centering
    \includegraphics[width=\linewidth]{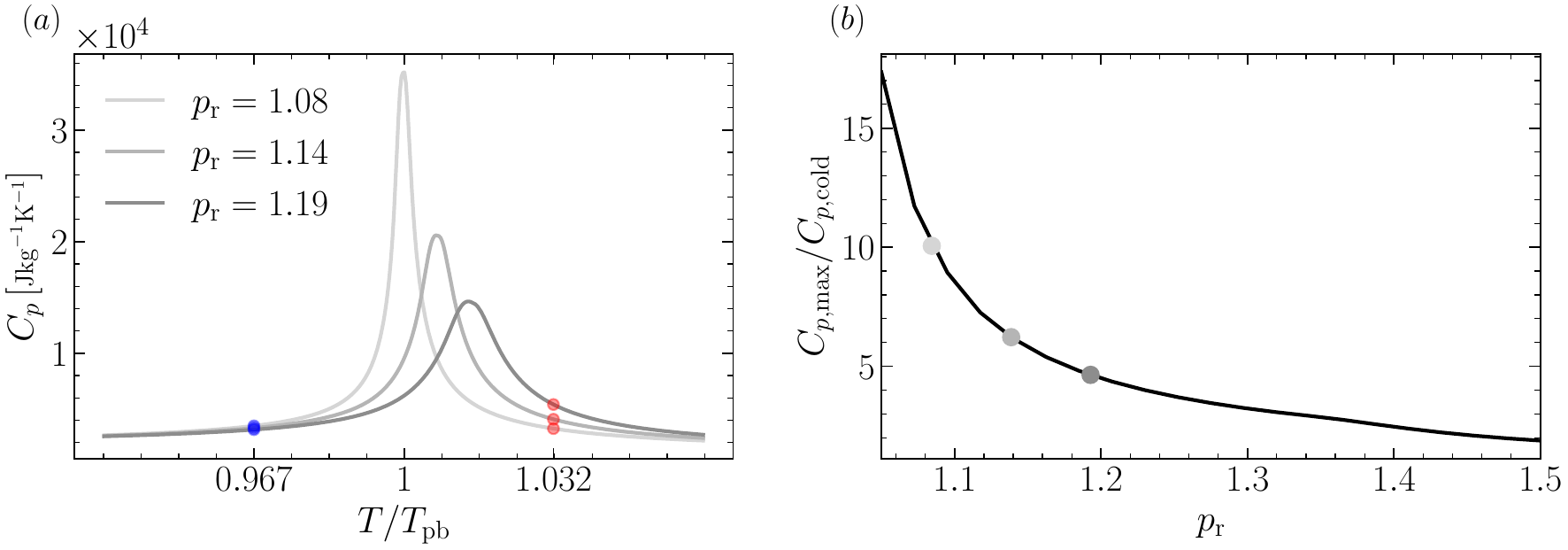}
    \caption{(\emph{a}) Variation of the specific isobaric heat capacity with temperature at different pressures. Here, $p_r = 1.08,\, 1.14,\, \text{and}\, 1.19$ correspond to $p_0 \approx 8,\, 8.4,\, \text{and}\, 8.8\, \mathrm{MPa}$ respectively. $T_{\mathrm{pb}}$ denotes the pseudo-boiling temperature at $8~\mathrm{MPa}$. The blue and red markers indicate the values at the cold and hot walls, respectively. (\emph{b}) Variation of the ratio of the maximum heat capacity to its value at the cold wall as a function of thermodynamic pressure. The grey markers indicate the ratios corresponding to the pressures shown in panel (\emph{a}). The strong pressure sensitivity of $C_p$ near the critical point highlights the importance of consistently evolving the thermodynamic pressure.}
    \label{fig:CpvsP}
\end{figure}

Figure~\ref{fig:CpvsP} further illustrates the variation of the ratio of the maximum heat capacity to the value at the cold wall as a function of thermodynamic pressure. Near the critical point, this quantity is highly sensitive to pressure, with relatively small pressure variations producing substantial changes in $C_p$. The ratio is close to $17$ near the critical point but decreases rapidly with increasing pressure, reaching approximately $2$ at $p_r = 1.5$. This pronounced pressure dependence highlights the importance of accurately evolving the thermodynamic pressure and, consequently, the thermodynamic state of the fluid. The local thermodynamic state determines the thermophysical properties, which in turn govern the wall heat transfer.

The initial temperature field is prescribed as a linear profile between $T_1$ and $T_2$. The resulting mean temperature $T = 307.7\ \mathrm{K}$ coincides with the pseudo-boiling temperature $T_{\mathrm{pb}}$\footnote{The pseudo-boiling point is defined as the temperature at which the specific heat capacity attains its maximum value at a given pressure, and marks the transition between liquid-like and gas-like fluid states. } at $p_{\mathrm{ref}} = 8\ \mathrm{MPa}$.
Using the thermodynamic-property tables, turbulent channel-flow simulations are performed in a computational domain of dimensions $L_x \times L_y \times L_z = 8 h\times 4h \times 2h$, with $2h = 2\times 10^{-4} \mathrm{m}$. The domain is discretized into $N_x \times N_y \times N_z = 512 \times 384 \times 384$ grid points. A constant pressure gradient drives the flow, such that the friction Reynolds number of $Re_{\tau,0} \approx 180$ is realized. Here, the friction Reynolds number is defined as $\rho_{\mathrm{pb}} u_{\tau,\mathrm{pb}} h/\mu_{\mathrm{pb}}$, where the subscript `$\mathrm{pb}$' corresponds to the pseudo-boiling point values. The simulation was carried out on two NVIDIA A100 GPUs for about $12$ hours to reach a fully developed state, and an additional $12$ hours to collect statistics.

\begin{figure}
    \centering
    \includegraphics[width=\linewidth]{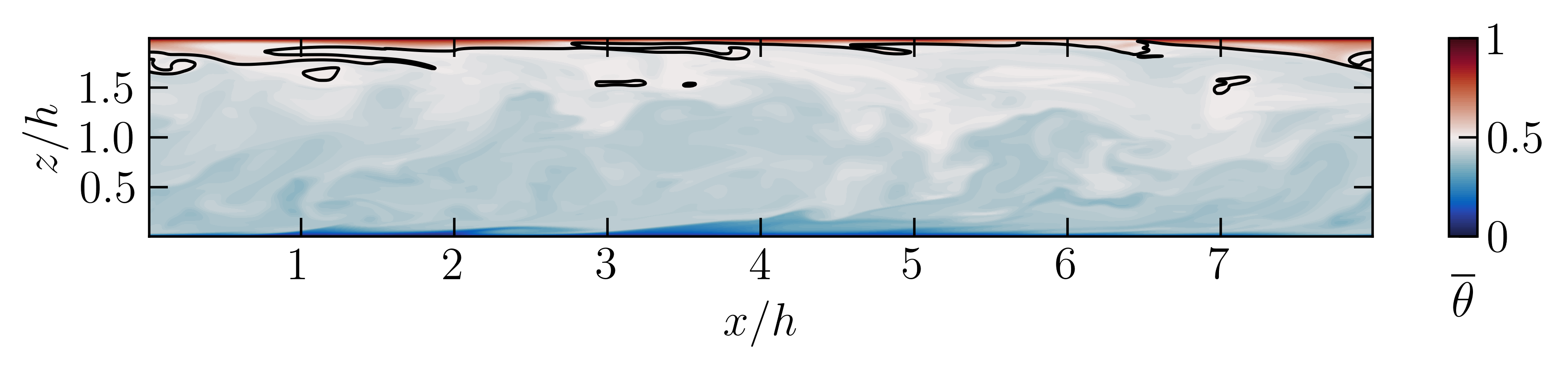}
    \caption{Instantaneous contours of temperature taken on the mid-spanwise plane for turbulent channel flow of transcritical $\mathrm{CO}_2$ at $p_0(0) = 8.4\ \mathrm{MPa}$ and $T_1 = 297.8\ \mathrm{K}$ ($\theta = 0$), $T_2\ = 317.8\ \mathrm{K}$ ($\theta = 1$). The pseudo-boiling point at $\Theta_\mathrm{pb}$, indicated by the black line, is shifted towards the hot wall. The steady-state thermodynamic pressure is $p_0 = 8\ \mathrm{MPa}$.}
    \label{fig:temp_contours_CO2}
\end{figure}

\begin{figure}
    \centering
    \begin{overpic}[width=\linewidth]{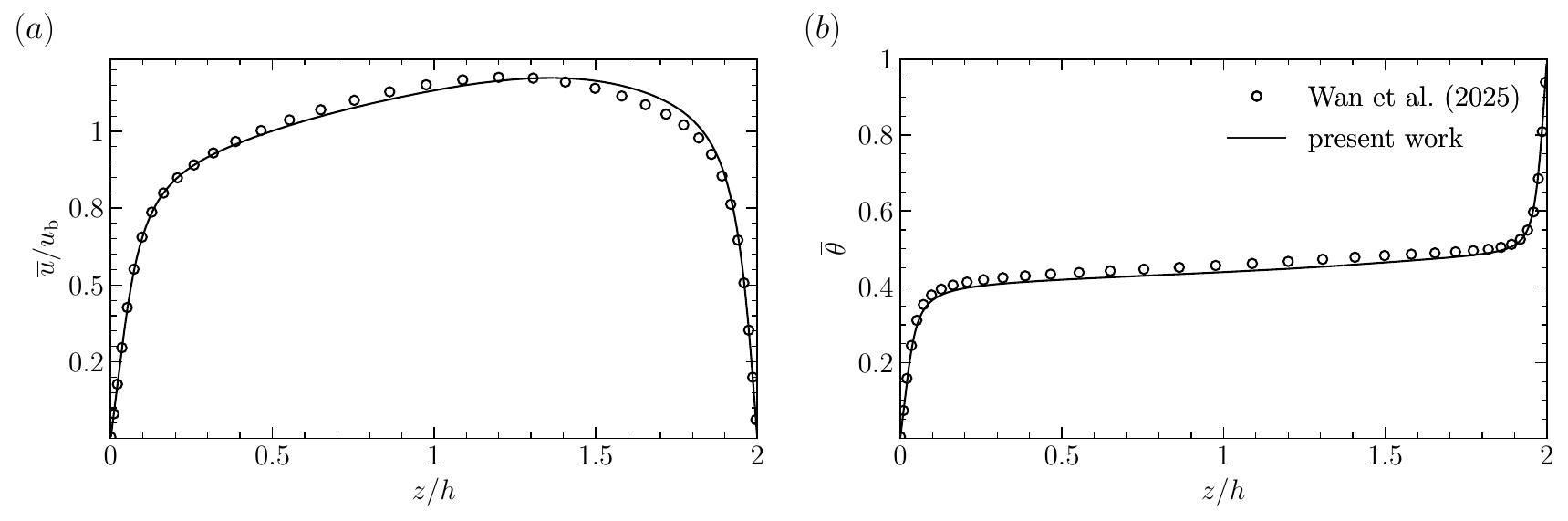}
        \put(81.4,23.6){\colorbox{white}{\normalsize Present work}}
    \end{overpic}
    \caption{(\emph{a}) Mean velocity profile and (\emph{b}) mean temperature profile for turbulent channel flow of transcritical $\mathrm{CO}_2$ at $p_0(0) = 8.4\ \mathrm{MPa}$ and $T_1 = 297.8\ \mathrm{K}$, $T_2\ = 317.8\ \mathrm{K}$. The mean velocity is normalized by the bulk velocity, while the temperature is normalized as $\theta = (\overline{T} - T_c)/(T_h - T_c)$, where the subscripts $c$ and $h$ correspond to the cold and hot walls, respectively. Present results are compared with the data of \citet{wan2025effects}.}
    \label{fig:vel_CO2}
\end{figure}

\begin{figure}
    \centering
    \includegraphics[width=.7\linewidth]{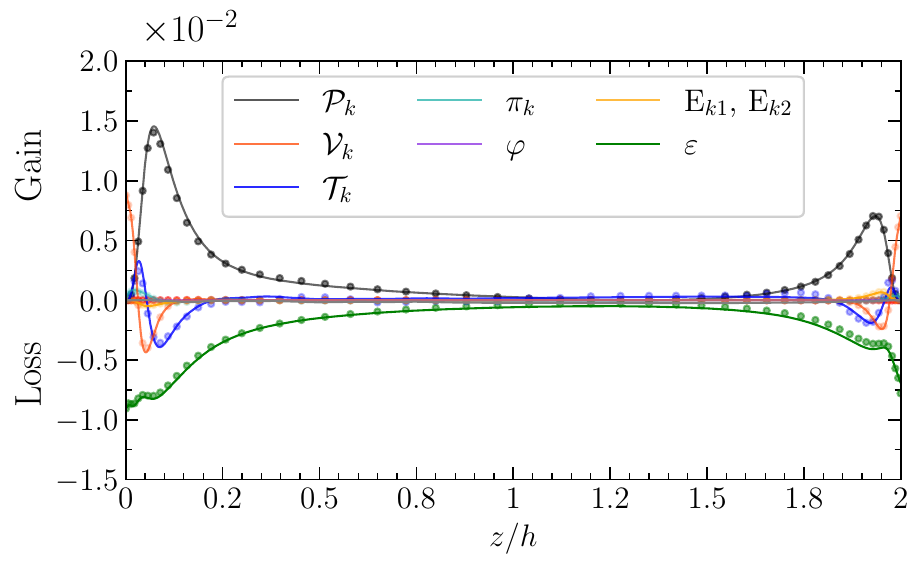}
    \caption{Comparison of the turbulent kinetic-energy budget terms for transcritical ($\mathrm{CO}_2$) channel flow at $p_0(0)=8.4~\mathrm{MPa}$, $T_1=297.8~\mathrm{K}$, and $T_2=317.8~\mathrm{K}$. Markers denote the reference DNS data of \citet{wan2025effects}, while solid lines correspond to the present results. The terms are based on Reynolds averages in the homogeneous directions and time, use the definitions in Eq.~\eqref{eq:tke_budget}, and are plotted in the nondimensional form supplied with the reference data; the ordinate multiplier is shown on the axis.}
    \label{fig:TKE_budget_CO2}
\end{figure}

Instantaneous contours of the temperature difference on the mid-spanwise plane of the fully developed turbulent flow of transcritical $\mathrm{CO}_2$ are presented in Figure~\ref{fig:temp_contours_CO2}. The higher value of thermal conductivity on the cold side results in smoother gradients than on the hot side. This also shifts the pseudo-boiling point towards the hot wall, as evidenced by the temperature contours. Furthermore, the mean velocity and mean temperature profiles obtained from the simulation are compared against \citet{wan2025effects} in Figure~\ref{fig:vel_CO2}. The statistics are averaged over a period of approximately 50 eddy turnover times after the flow has reached a statistically steady state. The present results show good agreement with the reference data. Further validation is provided through a comparison of the turbulent kinetic-energy (TKE) budget terms shown in Figure~\ref{fig:TKE_budget_CO2}. For low–Mach–number variable-property flows, we use $u_i = \tilde{u}_i + u_i''$, $\tilde{u}_i = \overline{\rho u_i}/\overline{\rho}$, and $\overline{\rho u_i''}=0$, with overbars denoting Reynolds averages.

The turbulent kinetic-energy transport equation is
\begin{equation}
\begin{split}
    \frac{\partial \overline{\rho} k}{\partial t} + {\frac{\partial \overline{\rho}\tilde{u}_j k}{\partial x_j }} & = \underbrace{- \overline{\rho \ u_j'' u_i'' }\frac{\partial \tilde{u_i}}{\partial x_j}}_{\mathcal{P}_k} - \\
    & \underbrace{\frac{\partial}{\partial x_j}\left(\frac{ \overline{\rho u_j'' u_i'' u_i''}}{2}\right)}_{\mathcal{T}_k} - \underbrace{\frac{\partial}{\partial x_j}\left(\overline{u_j''p_1'}\right)}_{\mathcal{\pi}_k} + \underbrace{\frac{\partial}{\partial x_j}\left( \overline{u_i''\tau_{ij}'}\right)}_{\mathcal{V}_k} -\\
    & \underbrace{\overline{u_i''}\frac{\partial \overline{p_1}}{\partial x_i}}_{\mathrm{E}_{k1}} + \underbrace{\overline{u_i''}\frac{\partial \overline{\tau}_{ij}}{\partial x_j}}_{\mathrm{E}_{k2}} + \underbrace{\overline{p_1'\frac{\partial u_i''}{\partial x_i}}}_{\varphi} - \underbrace{\overline{\tau_{ij}'\frac{\partial u_i''}{\partial x_j}}}_{\varepsilon},
\end{split}\label{eq:tke_budget}
\end{equation}
where $k=\widetilde{u_i''u_i''}/2$ is the turbulent kinetic energy and $p_1'=p_1-\overline{p_1}$ is the Reynolds fluctuation of the hydrodynamic pressure. The terms $\mathcal{P}_k$, $\mathcal{T}_k$, $\mathcal{\pi}_k$, and $\mathcal{V}_k$ denote the production, turbulent transport, pressure transport, and viscous transport, respectively. The term $\varphi$ represents pressure dilatation, while $\mathrm{E}_{k1}$ and $\mathrm{E}_{k2}$ are additional contributions arising from density fluctuations. Finally, $\varepsilon$ denotes the dissipation rate of turbulent kinetic energy. Figure~\ref{fig:TKE_budget_CO2} compares the individual budget terms with the DNS data of \citet{wan2025effects}. The half-channel height and bulk velocity are used as the normalizing scales. Very good agreement is observed for all terms, further demonstrating the ability of the proposed framework to accurately predict both the mean flow and turbulence statistics.

\section{Conclusion}\label{sec:conclusion}
A low-Mach number framework for the simulation of low-speed non-ideal-fluid flows with large temperature variations has been presented. While the formulation is applicable to both open and closed systems, one of the contributions of this work is an efficient mass-preserving treatment of thermodynamic-pressure evolution in closed systems. Because the equation of state is nonlinear, unlike the ideal-gas formulation, the thermodynamic pressure cannot be obtained explicitly; a Newton--Raphson procedure is therefore developed to determine it while enforcing global mass conservation. The resulting numerical algorithm employs a segregated strategy in which the thermodynamic state is advanced prior to the momentum equations, together with a pressure-splitting approach that enables efficient FFT-based solution of the pressure-correction equation. The framework is robust and flexible and can be applied to arbitrary pure fluids within a stable single-phase regime through coupling with tabulated thermodynamic data. The implementation is embedded in a distributed-memory, GPU-accelerated solver, enabling fast, large-scale simulations of turbulent flows of non-ideal fluids with strong property variations.

The implementation was verified using the method of manufactured solutions. To accommodate non-ideal (e.g., cubic) equations of state, the manufactured thermodynamic state was constructed from analytical density and thermodynamic-pressure fields rather than by prescribing temperature directly. The temperature field was then obtained from the equation of state. This enabled simultaneous verification of the nonlinear equation of state, thermodynamic-pressure evolution, and the velocity-divergence constraint. The verification study demonstrated the expected second-order accuracy of the solver.

The framework was subsequently validated against benchmark problems for ideal-gas flows, including the differentially heated cavity and turbulent channel flow, showing good agreement with reference data. The differentially heated cavity was also computed with a van der Waals fluid at a characteristic Mach number of $\mathcal{O}(10^{-7})$, a regime in which a fully compressible treatment is impractical, highlighting the advantage of the present low-Mach framework for non-ideal fluids. Finally, the framework was validated against turbulent channel flows at transcritical conditions. Comparisons with the DNS data of \citet{wan2025effects} showed very good agreement in both the mean velocity profiles and turbulent kinetic-energy budgets, demonstrating the capability of the proposed framework to accurately capture the coupled thermodynamic and hydrodynamic behavior of non-ideal fluids with strong property variations.

\section*{Acknowledgements}
The authors thank Prof.~C\'elio Fernandes from the University of Porto -- Faculty of Engineering, for insightful discussions, and Prof.~Xingjian Wang from Tsinghua University, for kindly sharing the data from \cite{wan2025effects} in digital form. This work was supported by the European Research Council (grant No.~\texttt{ERC-2019-CoG-864660}, \emph{Critical}), and NVIDIA Corporation (Academic Grant \emph{Turbulent forced convection beyond the Oberbeck-Boussinesq hypothesis}).

\appendix
\counterwithin{equation}{section}
\counterwithin{figure}{section}
\counterwithin{table}{section}

\section{Derivation of the formulation}\label{sec:LMDerivation}
The conservation equations of mass, momentum, and sensible enthalpy for a fully compressible fluid are \citep{poinsot2005theoretical}
\begin{subequations}\label{eq:fully_compressible_system}
\begin{equation}
    \frac{\partial \rho}{\partial t} + \frac{\partial}{\partial x_i} \left(\rho  u_i\right) = 0;
\end{equation}

\begin{equation}
    \frac{\partial (\rho u_j)}{\partial t} + \frac{\partial}{\partial x_i}\left(\rho u_i u_j\right) = -\frac{\partial p}{\partial x_j} + \frac{\partial \tau_{ij}}{\partial x_i} + \rho f_j;
\end{equation}

\begin{equation}\label{eq:app_enthalpy}
    \frac{\partial (\rho h)}{\partial t} + \frac{\partial}{\partial x_i}\left(\rho u_i h\right) = \frac{\mathrm{d}p}{\mathrm{d}t} - \frac{\partial q_i}{\partial x_i} + \tau_{ij}\frac{\partial u_i}{\partial x_j} + \mathcal{Q}.
\end{equation}
\end{subequations}
Here, $\rho$ is the density, $u_i$ the velocity component in the $x_i$ direction, $p$ the pressure, and $h$ the sensible enthalpy. The viscous stress tensor is defined as
\begin{equation}
    \tau_{ij} = \mu \left(\frac{\partial u_i}{ \partial x_j} + \frac{\partial u_j}{ \partial x_i} - \frac{2}{3}\ \frac{\partial u_k}{\partial x_k}\ \delta_{i,j}\right),
\end{equation} 
with $\mu$ being the dynamic viscosity. Here, $f_j$ is a body-force acceleration and $\mathcal{Q}$ is a volumetric heat source. The heat flux vector $q_i$ is related to temperature ($T$) gradients using Fourier's law as 
\begin{equation}
    q_i = -\lambda \frac{\partial T}{\partial x_i}, 
\end{equation}
where $\lambda$ is the thermal conductivity.

\paragraph{Non-dimensionalisation:} The dimensionless form of the governing equations is obtained by normalizing all variables using appropriate reference scales. The characteristic length and velocity scales are denoted by $\mathcal{L}$ and $\mathcal{U}$, respectively. The reference pressure and density are chosen as $\mathcal{P}$ and $\mathcal{\varrho}$. Based on these, other reference quantities can be defined. For instance, the reference time is $\mathtt{t}_0 = {\mathcal{L}}/{\mathcal{U}}$.

The reference temperature can be expressed in terms of the reference pressure and density as
\begin{equation}
    \mathcal{T} = \mathcal{T}(\mathcal{P}, \varrho) = \frac{1}{\mathcal{Z}\mathcal{R}} \frac{\mathcal{P}}{\varrho},
\end{equation}
where $\mathcal{R}$ is the specific gas constant (the universal gas constant divided by the molecular mass), and $\mathcal{Z}$ is the compressibility factor accounting for deviations from ideal gas behavior. 

The reference speed of sound ($\mathtt{a}_0$) is expressed in terms of thermodynamic quantities using the real gas isentropic exponent ($\gamma_{\mathrm{p\upsilon}}$), as demonstrated by \citet{Nederstigt2023GeneralisedThermodynamics}:
\begin{subequations}\label{eq:reference_sound_speed}
\begin{equation}
    \mathtt{a}_0^2 = \gamma_{\mathrm{p\upsilon}} \frac{\mathcal{P}}{\varrho} = \gamma_{\mathrm{p\upsilon}} \mathcal{Z}\mathcal{R}\mathcal{T},
\end{equation}
where
\begin{equation}\label{eq:isent_gamma}
    \gamma_{\mathrm{p\upsilon}} = \frac{1}{\mathcal{P}}\frac{C_p}{C_{\upsilon}}\frac{1}{\chi}.
\end{equation}
\end{subequations}

Furthermore, the reference enthalpy ($\mathtt{h}_0$) is determined by the thermodynamic state of the system and can be expressed as $\mathtt{h}_0(\mathcal{P},\varrho)$. In the subsequent derivation, $\mathtt{h}_0$ is taken to scale with the square of the speed of sound, the natural thermodynamic scale, and written as
\begin{equation}
    \mathtt{h}_0 \sim \mathtt{a}_0^2 = \gamma_{\mathrm{p\upsilon}} \frac{\mathcal{P}}{\varrho} = \gamma_{\mathrm{p\upsilon}} \mathcal{Z}\mathcal{R}\mathcal{T}.
\end{equation}

The non-dimensional numbers relevant to the dimensionless system are
\begin{equation}
    Re = \frac{\varrho\, \mathcal{U}\, \mathcal{L}}{\mu_0},
    \quad
    {Ma} = \frac{\mathcal{U}}{\mathtt{a}_0}.
\end{equation}
Here, $Re$ is the Reynolds number and $Ma$ is the Mach number.

Using these reference values, the variables are non-dimensionalized as
\begin{subequations}\label{eq:nondimensional_variables}
\begin{equation}
    \widetilde x_i = \frac{x_i}{\mathcal{L}}; 
    \quad
    \widetilde u_i = \frac{u_i}{\mathcal{U}};
    \quad
    \widetilde t = \cfrac{t}{\mathcal{L}/\mathcal{U}};
\end{equation}
\begin{equation}
    \widetilde p = \frac{p}{\mathcal{P}}; 
    \quad
    \widetilde \rho = \frac{\rho}{\varrho};
    \quad
    \widetilde h = \cfrac{h}{\mathtt{h}_0} = \frac{h}{\gamma_{\mathrm{p\upsilon}}\mathcal{P}/\varrho};
\end{equation}

\begin{equation}
    \widetilde \tau_{ij} = \frac{\tau_{ij}}{\mu_0\ \mathcal{U}/\mathcal{L}}; 
    \quad
    \widetilde q_i = \frac{q_i}{\lambda_0\, \mathcal{T}/\mathcal{L}}.
\end{equation}
\end{subequations}
The remaining source terms are scaled as $\widetilde f_i=f_i\mathcal{L}/\mathcal{U}^2$ and $\widetilde{\mathcal{Q}}=\mathcal{Q}\mathcal{L}/(\varrho\mathcal{U}\mathtt{h}_0)$. With these definitions, the dimensionless system is
\begin{subequations}\label{eq:dimensionless_system}
\begin{equation}
    \frac{\partial \widetilde \rho}{\partial \widetilde t} + \frac{\partial}{\partial \widetilde x_i} \left(\widetilde \rho\, \widetilde u_i\right) = 0;
\end{equation}

\begin{equation}
    \frac{\partial (\widetilde \rho \, \widetilde u_i)}{\partial \widetilde t} + \frac{\partial}{\partial \widetilde x_j}\left(\widetilde \rho\, \widetilde u_j\widetilde u_i\right) = -\frac{\partial \widetilde p}{\partial \widetilde x_i}\, \cfrac{1}{\gamma_{\mathrm{p\upsilon}}\, Ma^2} + \frac{\partial \widetilde \tau_{ij}}{\partial\widetilde x_j}\, \frac{1}{Re} + \widetilde\rho\,\widetilde f_i;
\end{equation}

\begin{equation}
    \cfrac{\partial (\widetilde \rho\, \widetilde h)}{\partial \widetilde t} + \cfrac{\partial}{\partial \widetilde x_i}\left(\widetilde\rho\, \widetilde u_i \widetilde h\right) = \cfrac{1}{\gamma_{\mathrm{p\upsilon}}}\cfrac{\mathrm{d}\widetilde p}{\mathrm{d}\widetilde t} - \frac{1}{Re\,\widehat{Pr}}\cfrac{\partial \widetilde q_i}{\partial \widetilde x_i} + \frac{{Ma}^2}{Re}\, \widetilde\tau_{ij}\cfrac{\partial \widetilde u_i}{\partial \widetilde x_j} + \widetilde{\mathcal{Q}}.
\end{equation}
\end{subequations}
The non-dimensional group $\widehat{Pr}=\mu_0\gamma_{\mathrm{p\upsilon}}\mathcal{Z}\mathcal{R}/\lambda_0$ is analogous to the Prandtl number for ideal gases\footnote{For an ideal gas, $\mathcal{Z}=1$ and $C_p=\gamma R/(\gamma-1)$ yield $\widehat{Pr}=(\gamma-1)Pr$.}, up to the factor $\gamma-1$ \citep{sirignano2018compressible}. This difference arises from the choice of enthalpy reference scale adopted in the present non-dimensionalization, which differs from the conventional ideal-gas scaling based on the specific-enthalpy scale $\mathtt{h}_0=C_{p0}\mathcal{T}$.

The dependent variables are expressed as asymptotic expansions in the parameter $\gamma_{\mathrm{p\upsilon}}\,Ma^2$ as
\begin{subequations}\label{eq:asymptotic_expansions}
\begin{equation}
    \widetilde p = \widetilde{p}_0 + (\gamma_{\mathrm{p\upsilon}}\, Ma^2) \ \widetilde p_1 + (\gamma_{\mathrm{p\upsilon}}\, Ma^2)^2 \ \widetilde p_2 + \cdots;
\end{equation}
\begin{equation}
        \widetilde \rho = \widetilde{\rho}_0 + (\gamma_{\mathrm{p\upsilon}}\, Ma^2) \ \widetilde \rho_1 + (\gamma_{\mathrm{p\upsilon}}\, Ma^2)^2 \ \widetilde \rho_2 + \cdots;
\end{equation}
\begin{equation}
        \widetilde u = \widetilde{u}_0 + (\gamma_{\mathrm{p\upsilon}}\, Ma^2) \ \widetilde u_1 + (\gamma_{\mathrm{p\upsilon}}\, Ma^2)^2 \ \widetilde u_2 + \cdots,
\end{equation}
\end{subequations}
with an analogous expansion for the temperature.

In the low–Mach-number limit, $\gamma_{\mathrm{p\upsilon}}\, Ma^2 \ll 1$, higher-order terms become progressively less significant. Substituting these expansions into the dimensionless governing equations and collecting terms of equal order in $\gamma_{\mathrm{p\upsilon}}Ma^2$ yields the leading-order system. Because the leading-order pressure is spatially uniform, $\partial\widetilde p_0/\partial\widetilde x_i=0$, its material derivative reduces exactly to the ordinary time derivative, $\mathrm{D}\widetilde p_0/\mathrm{D}\widetilde t=\mathrm{d}\widetilde p_0/\mathrm{d}\widetilde t$.

At zeroth order, $\mathcal{O}\left((\gamma_{\mathrm{p\upsilon}} Ma^2)^0\right)$, the governing equations reduce to
\begin{subequations}\label{eq:zeroth_order_system}
\begin{equation}
    \frac{\partial \widetilde \rho_0}{\partial \widetilde t} + \frac{\partial}{\partial \widetilde x_j}\left(\widetilde \rho_0 \widetilde u_{j0}\right) = 0;
\end{equation}
\begin{equation}
    \frac{\partial \widetilde p_0}{\partial \widetilde x_j} = 0;
\end{equation}
\begin{equation}
    \cfrac{\partial (\widetilde \rho_0\, \widetilde h_0)}{\partial \widetilde t} + \cfrac{\partial}{\partial \widetilde x_j}\left(\widetilde\rho_0\, \widetilde {u_j}_0 \widetilde h_0\right) = \cfrac{1}{\gamma_{\mathrm{p\upsilon}}}\cfrac{\mathrm{d}\widetilde p_0}{\mathrm{d}\widetilde t} - \cfrac{1}{Re\,\widehat{Pr}} \cfrac{\partial \widetilde q_j}{\partial \widetilde x_j} +\widetilde{\mathcal{Q}}.
\end{equation}
\end{subequations}

The zeroth-order pressure $\widetilde p_0$ does not contribute to momentum transport and is spatially uniform. It describes the thermodynamic state of the system and is therefore referred to as the \emph{thermodynamic pressure}. Momentum transport arises from perturbations in first-order pressure ($\widetilde p_1$), also referred to as the \emph{hydrodynamic pressure}.

At first order, $\mathcal{O}\left((\gamma_{\mathrm{p\upsilon}} Ma^2)^1\right)$, the momentum equation becomes
\begin{equation}
    \frac{\partial (\widetilde \rho_0 \, \widetilde {u_i}_0)}{\partial \widetilde t} + \frac{\partial}{\partial \widetilde x_j}\left(\widetilde \rho_0\, \widetilde {{u}_j}_0\widetilde {{u}_i}_0\right) = -\frac{\partial \widetilde p_1}{\partial \widetilde x_i}\  + \frac{1}{Re}\, \frac{\partial \widetilde {\tau_{ij}}_0}{\partial\widetilde x_j}  + \widetilde\rho_0\,\widetilde f_i.
\end{equation}

Since all dependent variables enter the governing equations at zeroth order, with the exception of the pressure, the subscript $0$ is omitted for all variables except the pressure. Under this convention, the dimensional low–Mach number governing equations can be written as follows.
\begin{subequations}\label{eq:dimensional_low_mach_system}
\begin{equation}\label{eq:mass1}
    \frac{\partial \rho}{\partial t} + \frac{\partial \rho u_i}{\partial x_i} = 0;
\end{equation} 

\begin{equation}\label{eq:momentum0}
    \frac{\partial p_0}{\partial x_i} = 0;
\end{equation}

\begin{equation}\label{eq:momentum1}
    \frac{\partial (\rho u_i)}{\partial t} + \frac{\partial }{\partial x_j}({\rho u_i u_j}) = -\frac{\partial p_1}{\partial x_i} + \frac{\partial \tau_{ij}}{\partial x_j} + \rho f_i;
\end{equation}

\begin{equation}\label{eq:enthalpy1}
     \frac{ \partial (\rho h) }{\partial t} + \frac{\partial }{\partial x_i}({\rho u_i h}) = \frac{\partial }{\partial x_i}\left(\lambda \frac{\partial T}{\partial x_i}\right)  + \dot{\mathcal{Q}}  + \frac{\mathrm{d} p_0}{\mathrm{d} t}.
\end{equation}
\end{subequations}

Note that the viscous dissipation term in Eq.~\eqref{eq:app_enthalpy} vanishes in the low-Mach-number formulation, as it is of order $\mathcal{O}\gamma_{\mathrm{p\upsilon}}Ma^2$.

\section{Recasting the enthalpy transport equation}\label{sec:enth2temp}
The enthalpy transport equation is
\begin{equation}\label{eq:sstep0}
     \frac{ \partial (\rho h) }{\partial t} + \frac{\partial }{\partial x_i}({\rho u_i h}) = \frac{\partial }{\partial x_i}\left(\lambda \frac{\partial T}{\partial x_i}\right)  + \dot{\mathcal{Q}}  + \frac{\mathrm{d} p_0}{\mathrm{d} t}.
\end{equation}
It is recast in terms of temperature, which is the primary thermodynamic variable advanced in the numerical algorithm. The enthalpy differential and the definition of $C_p$ give
\begin{subequations}\label{eq:enthalpy_differential}
\begin{equation}\label{eq:enthalpy_total_differential}
    \mathrm{d}h = \left. \frac{\partial h}{\partial T}\right|_{p} \mathrm{d} T + \left. \frac{\partial h}{\partial p}\right|_{T} \mathrm{d} p;
\end{equation}
\begin{equation}\label{eq:sstep1}
    \mathrm{d}h =C_p\ \mathrm{d} T + \left. \frac{\partial h}{\partial p}\right|_{T} \mathrm{d} p.
\end{equation}
\end{subequations}

To simplify the second term on the right-hand side of Eq.~\eqref{eq:sstep1}, we use the thermodynamic identity and a Maxwell relation:
\begin{subequations}\label{eq:enthalpy_pressure_derivative}
\begin{equation}\label{eq:gibbs_enthalpy_relation}
    \mathrm{d} h = T\mathrm{d}s + v\mathrm{d}p;
\end{equation}
\begin{equation}\label{eq:enthalpy_pressure_intermediate}
    \left. \frac{\partial h}{\partial p}\right|_{T} = T \left. \frac{\partial s}{\partial p}\right|_{T} + v;
\end{equation}
\begin{equation}\label{eq:enthalpy_pressure_maxwell}
    \left. \frac{\partial h}{\partial p}\right|_{T} = -T \left. \frac{\partial v}{\partial T}\right|_{p} + v.
\end{equation}
\end{subequations}

Substituting this result into Eq.~\eqref{eq:sstep1}, with $v=1/\rho$ and the definition of $\beta$, gives
\begin{equation}\label{eq:sstep2}
    \begin{aligned}
    \mathrm{d}h
        &=C_p\ \mathrm{d} T + \left(-T \left. \frac{\partial v}{\partial T}\right|_{p} + v\right) \mathrm{d} p \\
        &= C_p\ \mathrm{d} T + \left(-\beta \frac{T}{\rho} + \frac{1}{\rho}\right) \mathrm{d} p.
    \end{aligned}
\end{equation}

Taking the material derivative and multiplying by $\rho$ yields
\begin{equation}\label{eq:sstep4}
    \rho \frac{\mathrm{D}h}{\mathrm{D}t} = \rho C_p\ \frac{\mathrm{D}T}{\mathrm{D}t} + \left(1 - \beta T \right) \frac{\mathrm{D}p}{\mathrm{D}t}.
\end{equation}
where the $LHS$ of Eq.\eqref{eq:sstep4} is same as the $LHS$ of Eq.\eqref{eq:sstep0}.

In the low--Mach--number system, the thermodynamic pressure entering this identity is the spatially uniform $p_0(t)$. Hence, $\mathrm{D}p_0/\mathrm{D}t=\mathrm{d}p_0/\mathrm{d}t$, and substituting Eq.~\eqref{eq:sstep4} into the enthalpy transport equation yields
\begin{equation}\label{eq:rg_enthalp}
     \rho C_p\ \frac{\mathrm{D}T}{\mathrm{D}t} = -\frac{\partial q_i}{\partial x_i}  + \dot{\mathcal{Q}}  + \beta T \frac{\mathrm{d} p_0}{\mathrm{d} t}.
\end{equation}

\section{Heat capacities for the Van der Waals fluid}\label{sec:cpvdw}
The isobaric heat capacity is defined by its enthalpy derivative and, for a generic gas, may be expressed as
\begin{subequations}\label{eq:general_heat_capacities}
\begin{equation}\label{eq:cp_definition}
    C_p = \left.\frac{\partial h}{\partial T}\right|_p;
\end{equation}
\begin{equation}\label{eq:cp_general_relation}
    C_p = C_{\upsilon} + \frac{T}{\rho^2} \left.\frac{\partial\rho}{\partial p}\right|_T  \left(\left.\frac{\partial p}{\partial T}\right|_{\rho}\right)^2,
\end{equation}
\end{subequations}
where $C_{\upsilon}$ is the isochoric heat capacity. Note that the pressure corresponds to thermodynamic pressure, $p_0$.

For a Van der Waals fluid, the isochoric heat capacity is determined by the number of active degrees of freedom, $f$. The reduced heat capacities are
\begin{subequations}\label{eq:vdw_heat_capacities}
\begin{equation}\label{eq:vdw_cv}
    C_{\upsilon,r} = \frac{C_{\upsilon}}{R} = \frac{f}{2};
\end{equation}
\begin{equation}\label{eq:vdw_cp}
    C_{p,r} = C_{\upsilon,r} + \left[1 - \frac{\rho_r(3 - \rho_r)^2}{4T_r}\right]^{-1}.
\end{equation}
\end{subequations}
Here, $R$ denotes the specific gas constant, and the subscript `$r$' corresponds to reduced values. In the present work, $\mathrm{CO}_2$ is considered, for which $f = 9$ \citep{boldini2025direct}.

\bibliography{literature}
\end{document}